\documentclass[useAMS,usenatbib]{mn2e}
\usepackage{graphicx}
\usepackage[varg]{txfonts}
\usepackage{natbib}
\usepackage{color}

\title[AGB evolution in bulge stars]{Understanding AGB evolution in Galactic
bulge stars from high-resolution infrared spectroscopy}
\author[Uttenthaler et al.]
{S.~Uttenthaler,$^{1}$\thanks{stefan.uttenthaler@univie.ac.at (SU);
    joris.blommaert@vub.ac.be (JADLB); thomas.lebzelter@univie.ac.at (TL)}
J.~A.~D.~L.~Blommaert,$^{2,3}$ P.~R.~Wood,$^{4}$ T.~Lebzelter,$^{1}$
 B.~Aringer,$^{5,1}$
\newauthor
M.~Schultheis,$^{6}$ N.~Ryde$^{7}$\\
$^{1}$University of Vienna, Department of Astrophysics, T\"urkenschanzstra\ss e 17, 1180 Vienna, Austria\\
$^{2}$Astronomy and Astrophysics Research Group, Department of Physics and Astrophysics, Vrije Universiteit Brussel, Pleinlaan 2, 1050 Brussels, Belgium\\
$^{3}$Instituut voor Sterrenkunde, KU Leuven, Celestijnenlaan 200D, 3001 Leuven, Belgium\\
$^{4}$Research School of Astronomy and Astrophysics, Australian National University, Cotter Road, Weston Creek ACT 2611, Australia\\ 
$^{5}$Department of Physics and Astronomy G.\ Galilei, University of Padova, Vicolo dell'Osservatorio 3, I-35122 Padova, Italy\\
$^{6}$Laboratoire Lagrange (UMR7293), Universit\'{e} de Nice Sophia Antipolis, CNRS, Observatoire de la C\^{o}te d’Azur, BP 4229, 06304 Nice Cedex 4, France\\
$^{7}$Lund Observatory, Box 43, 221 00 Lund, Sweden}
\begin{document}

\date{Accepted 2015 May 6. Received 2015 April 17; in original form 2015 March 8}

\pagerange{\pageref{firstpage}--\pageref{lastpage}} \pubyear{201x}

\maketitle

\label{firstpage}

\begin{abstract}
An analysis of high-resolution near-infrared spectra of a sample of 45
asymptotic giant branch (AGB) stars towards the Galactic bulge is presented. The
sample consists of two subsamples, a larger one in the inner and intermediate
bulge, and a smaller one in the outer bulge. The data are analysed with the help
of hydrostatic model atmospheres and spectral synthesis. We derive the radial
velocity of all stars, and the atmospheric chemical mix ([Fe/H], C/O,
$^{12}$C/$^{13}$C, Al, Si, Ti, and Y) where possible. Our ability to model the
spectra is mainly limited by the (in)completeness of atomic and molecular line
lists, at least for temperatures down to $T_{\rm eff}\approx3100$\,K. We find that
the subsample in the inner and intermediate bulge is quite homogeneous, with a
slightly sub-solar mean metallicity and only few stars with super-solar
metallicity, in agreement with previous studies of non-variable M-type giants in
the bulge. All sample stars are oxygen-rich, C/O$<$1.0. The C/O and carbon
isotopic ratios suggest that third dredge-up (3DUP) is absent among the sample
stars, except for two stars in the outer bulge that are known to contain
technetium. These stars are also more metal-poor than the stars in the
intermediate or inner bulge. Current stellar masses are determined from linear
pulsation models. The masses, metallicities and 3DUP behaviour are compared to
AGB evolutionary models. We conclude that these models are partly in conflict
with our observations. Furthermore, we conclude that the stars in the inner and
intermediate bulge belong to a more metal-rich population that follows bar-like
kinematics, whereas the stars in the outer bulge belong to the metal-poor,
spheroidal bulge population.
\end{abstract}


\section{Introduction}\label{intro}

The Galactic bulge is a complex structure in the central regions of the Milky
Way Galaxy. Its stellar populations have been studied in some detail in the
recent years. Stellar types that have been investigated by high-resolution
spectroscopy in the past range from microlensed dwarf and subgiant stars
\citep[][ and references therein]{Bens13}, to K-type giants
\citep[e.g.][]{Bab10,Hill11,Utt12,Ness13}, to M-type giants
\citep[][ and references therein]{Rich12}. Also planetary nebulae (PNe) in the
bulge have been studied spectroscopically \citep{Mac99,Chi09,Guz11}.

A group of bulge stars that has so far not been studied in detail by
high-resolution spectroscopy consist of the asymptotic giant branch (AGB) stars.
AGB stars represent the last stage of stellar evolution where the energy output
is dominated by nuclear burning. This stage is characterised by strong
pulsations of the envelope that enhance the stellar wind, which is accelerated
by radiation pressure acting on dust particles formed in the outflowing
material. Eventually, the stellar evolution is terminated by the high mass loss
and the star may enter the PN phase. In the deep interior of the star,
nucleosynthetic processes take place that produce carbon ($^{12}$C) and heavy
elements. These may subsequently be brought to the surface by a deep mixing
event called the third dredge-up (3DUP). If enough C is added to the atmosphere,
the abundance ratio of C to O (by number of atoms) will exceed 1.0 and the star
will become a carbon star (C-star). The formation of C-stars depends on the
initial mass and metallicity: A star may turn into a C-star only if it is
massive enough to undergo enough 3DUP events, where the lower mass limit is a
function of metallicity, being lower for lower metallicity \citep{Mar13,Kar14}.
The lack of C-stars in the inner regions of M31 \citep{Boy13} suggests that
there could be a metallicity threshold above which the formation of C-stars is
inhibited altogether.

AGB stars in the Galactic bulge are interesting for a number of reasons. First
of all, it has been shown that intrinsic carbon stars, i.e.\ stars that owe
their enhancement of C to internal nucleosynthesis and 3DUP and not to binary
mass transfer, are widely absent in the Galactic bulge
\citep{Azz88,BT89,TR91,Ng97,SchultheDiss}. This would mean that either the mass
of bulge stars is too low or the metallicity too high to undergo (sufficient)
3DUP to form C-stars. However, it is known that oxygen-rich AGB stars in the
outer Galactic bulge show absorption lines of technetium (Tc), a clear indicator
of recent or ongoing 3DUP \citep{Utt07}. This suggests that 3DUP does happen in
bulge stars, but probably not enough to form C-stars. An exception may be the
first carbon Mira in the direction of the Galactic bulge that was recently
presented by \citet{Mis13}. It is not clear if that star is indeed a member of
the Galactic bulge\footnote{It was reported by \citet{CW02} that the Galactic
bar is traced by luminous, very red (in $J-K$) stars. These were suspected by
Cole \& Weinberg to be infrared carbon stars. However, these objects lack
spectroscopic confirmation of their C-star nature.}.

Samples of non-variable M-type giants, the precursors of variable AGB stars,
have been studied by \citet{RO05}, \citet{Rich07}, and \citet{Rich12}. One of
their main results is that the mean metallicity of M-giants in the bulge is
slightly sub-solar, and that stars with super-solar metallicity are not common.
On the other hand, it was concluded by \citet{Bue13} that PNe in the Galactic
bulge, the evolutionary successors of AGB stars, have a relatively high, i.e.\
on average super-solar, metallicity. Yet again, \citet{Utt12} suggests that AGB
stars in the outer bulge predominantly descend from the {\it metal-poor}
population ($[{\rm Fe}/{\rm H}]\approx-0.6$), not from the metal-rich
population ($[{\rm Fe}/{\rm H}]\approx+0.3$) of the bulge. This latter result
was derived from the fact that the radial velocity dispersion of AGB stars and
also of PNe in the outer bulge is similar to that of the metal-poor population,
which differs significantly from that of the metal-rich population.

Finally, \citet{PC09} and \citet{Guz11} showed that many Galactic bulge PNe
exhibit evidence of mixed chemistry with emission from both silicate dust and
polycyclic aromatic hydrocarbons (PAHs), the latter being a signature of C-rich
chemistry. Such signatures are difficult to explain in an environment with
(assumed) low C/O ratio, which required \citet{Guz11} to invoke hydrocarbon
chemistry in an UV-irradiated, dense torus around the central stars of the PNe.
In their chemical model, the formation of PAHs would be considerably eased if
the C/O ratio was increased from its initial value by at least a few 3DUP
events.

These considerations make Galactic bulge AGB stars interesting targets for
further investigation. A determination of the metallicity and mass of AGB stars
would help in understanding the lack of intrinsic C-stars in the bulge. In
particular, a metallicity determination would shed light on the discrepancies
in the metallicities of non-variable M-type giants and PNe in the bulge. A
measurement of the C/O ratio in bulge AGB stars would provide observational
constraints to explanations for the mixed-chemistry phenomenon in Galactic bulge
PNe.

In this paper we present near-IR spectroscopic observations of such a sample of
stars and the chemical composition derived from them. The near-IR range has a
number of advantages over the optical spectral range for studies of bulge red
giants, most of all the lower interstellar extinction, less crowding of lines,
and smaller effects of departures from local thermodynamic equilibrium
\citep[see][ for a detailed discussion]{Ryde09}. Also 3D effects are smaller in
the near-IR than in the optical range \citep{Kuc13}. Thus, with these data,
abundance determinations are done that would not be possible to obtain from
optical spectra, as used e.g.\ in \citet{Utt07}.

The paper is organised as follows: The sample selection, observations, and data
reduction are presented in Sect.~\ref{sel_obs_red}, the analysis of that data is
described in Sect.~\ref{analysis}; the results and their discussion are to be
found in Sect.~\ref{res_disc}; and conclusions are drawn in
Sect.~\ref{conclusio}.

\section{Sample selection, observations, and data reduction}\label{sel_obs_red}

In order to study the dust formation in the circumstellar environment of AGB
stars, a mid-infrared spectroscopic survey with the IRS spectrograph
\citep{Hou04} on board the {\it Spitzer} space telescope was performed on a
sample of Galactic bulge AGB stars
\citep[Spitzer Proposal ID \#3167, PI J.\ Blommaert;][]{Blo07,Vanh07,Gol14}. The
targets for that survey were selected from the ISOGAL survey \citep{Omo03} in
intermediate bulge fields ($|b|=1\degr$ to $4\degr$, corresponding distance from
the mid-plane $\sim140-560$\,pc at $R_0=8.0$\,kpc). The $(K_{\rm S}-[15])$
colour, where [15] is the 15\,$\mu$m band of the ISO satellite, is an estimate
of the dust mass-loss rate of the stars \citep{Ojha03,Blo06} and was used to
ensure that the full range of mass-loss rates, starting from the onset of (dust)
mass loss up to the mass-loss rates of the so-called super-winds of OH/IR stars
($10^{-4} M_{\sun}/{\rm yr}$) was included in the sample.

We selected from this {\it Spitzer} IRS sample those stars for high-resolution
near-IR observations for which we expected to be able to model their near-IR
spectra, i.e.\ OH/IR stars were excluded. Nevertheless, four large-amplitude
variables (LAVs) were observed. Following the recommendations of \citet{McS07},
these four LAVs were selected such that they would be near minimum light at the
time of the observing run. The initial goal of the observations was to determine
the C/O and $^{12}$C/$^{13}$C ratios of the stars as an ``evolutionary clock''
along the AGB as a result of repeated 3DUP. The observations were carried out
in visitor mode (observer: J.\ Blommaert) with the CRyogenic high-resolution
InfraRed Echelle Spectrograph \citep[CRIRES][]{HUK04} in three nights between
09 and 12 June 2008. The target stars themselves were used as guide star for
the wave-front sensor of the adaptive optics system. An entrance slit width of
0.4\arcsec was chosen, which results in a nominal resolving power of
$R=\lambda/\Delta\lambda=50\,000$.

Along with the 37 bulge AGB stars from the Spitzer sample, spectra were also
obtained of a few comparison stars that had been studied previously with
high-resolution near-IR spectroscopy by other authors. These include
BMB~78 studied by \citet{CS06}, BD$-$012971, which was studied by \citet{RO05},
as well as BMB~289, which was investigated by both these works. Furthermore,
spectra of five K/M-type field giants were obtained; the analysis and results
for the field stars are reported in the Appendix.

Also included in the present paper is an analysis of CRIRES spectra of eight AGB
stars in the outer bulge, the so-called Plaut stars \citep{Plaut}. They are much
further from the Galactic mid-plane than the Spitzer stars, at $b\sim-10\degr$
or $\sim1.4$\,kpc. These stars were selected from the sample discussed in
\citet{Utt07} and include all four stars that were found to contain Tc in their
atmosphere, as well as four Tc-poor stars. These observations were done either
in the science verification runs (August 2006) or in a regular service mode
observing programme (Prog.\ ID 383.D-0685(A), September 2009).

All stars were observed at least in one setting in the H-band (grating order 36)
and in the K-band (grating order 24). The H-band spectra cover approximately the
range $\lambda_{\rm vac}\simeq1543$ to 1561\,nm\footnote{All wavelengths in this
paper are expressed as nm {\em in vacuum}.}, with a gap of $\sim2$\,nm between
the detectors. Due to contamination from neighbouring orders, only data from
detector chips 2 and 3 are considered here. The H-band spectra are dominated by
the $^{12}$CO \mbox{3-0} band head, a number of OH and CN lines, as well as
atomic lines (Fe, Ni, Si, Ti, S). The K-band spectra in grating order 24 cover
the range $\lambda_{\rm vac}\simeq2355$ to 2407\,nm and are dominated by CO
$\Delta\nu=2$ lines.

Sixteen of the science targets and one of the comparison stars were observed in
two additional settings in the K-band to also have lines available of atomic
species that are relevant for the dust formation. These spectra in the grating
orders 27 and 26 cover the range $\lambda_{\rm vac}\simeq2087$ to 2134\,nm and
2174 -- 2222\,nm, respectively. Besides being cluttered with abundant CN lines,
these spectra contain, amongst others, lines of Na, Mg, Al, Si, Sc, Ti, and Fe.
The Na lines were targeted to check a possible correlation between Na line
equivalent width measured from low-resolution spectra and dust mineralogy in the
circumstellar envelope. This correlation was not reproduced with the
high-resolution spectra.

A basic reduction of the data was done with the standard ESO CRIRES pipeline,
version 1.7.0. Further processing, in particular wavelength calibration and
telluric correction, was done with custom-made IDL tools. The H-band spectra
were wavelength-calibrated manually using the ThAr lamp exposures taken at day
time. Due to the lack of suitable ThAr lines in the K-band settings, these
spectra were wavelength-calibrated using the numerous telluric features
imprinted in them.

Before dividing the science target spectrum by the telluric standard star
spectrum for telluric correction, both spectra were re-binned to a common
wavelength vector and the airmass of the telluric spectrum was adjusted to that
of the corresponding science spectrum according to the Lambert-Beer law:
\[
f_{\rm std,adj}=\exp\left[\ln\left(f_{\rm std}\right)\frac{X_{\rm sci}}{X_{\rm std}}\right],
\]
with $f_{\rm std}$ the flux of the telluric standard star spectrum
continuum-normalised to 1.0, and $X_{\rm{sci}}$ and $X_{\rm{std}}$ the airmass of
the science target and of the telluric standard star, respectively. The science
spectrum was then divided by the adjusted standard star spectrum. This careful
procedure in the telluric correction yielded very satisfactory results. Weak
residuals from the division were only discernible in the cores of the strongest
telluric lines. No telluric correction was applied to the H-band spectra because
the wavelength range under consideration is essentially free from telluric
absorption. The achieved signal-to-noise ratio (S/N) per pixel of the reduced
one-dimensional spectra is relatively high ($\sim100$) for most stars, only for
a few stars it is somewhat low (of the order of 30). The S/N is thus not the
limiting factor in the analysis of most stars.

Particular care was taken in the final step of data reduction, the continuum
normalisation. The prime method used in this step was to identify narrow
pieces of spectrum (a few 0.01 up to 0.10\,nm) that are basically unaffected by
line absorption in as many of the sample stars as possible. The mean flux of the
30\% brightest pixels in these ranges was set to define the con\-ti\-nu\-um
(flux=1.0). This continuum placement was always checked in comparison with
synthetic spectra. Figure~\ref{example_fit} provides an example of how well the
continuum in observed and synthetic spectra are matched. In some cases, in
par\-ti\-cu\-lar for the coolest and most strongly variable stars, even these
narrow pieces of spectrum were affected by line absorption. In these cases, the
continuum was placed manually, mostly by dividing it by its maximum flux, except
in cases that were obviously affected by bad pixels, residuals of the telluric
correction, or emission components. This method was necessary anyway in the
K-band spectra of grating order 24 because this spectral region is heavily
cluttered with CO $\Delta\nu=2$ lines.

\section{Analysis}\label{analysis}

\subsection{Stellar parameter determination}\label{stellar_params}

Before determining elemental abundances from the high-resolution spectra using
spectral synthesis techniques, the main physical parameters of our sample stars
need to be determined. We derived those stellar parameters from low-resolution
optical spectra, broad band photometry, and the pulsation periods of the stars.
These data will be described in more detail in a separate paper (Vanhollebeke et
al., in preparation).

\subsubsection{Pulsation period}\label{puls_period}

The pulsation periods of the sample stars are an important ingredient for the
stellar parameter determination. K-band photometry of the Spitzer sub-sample
spanning $\sim820$ days was obtained by one of us (PW) and analysed by
\citet{Vanh07}, who also derived pulsation periods where possible. Furthermore,
the VizieR\footnote{\tt http://vizier.u-strasbg.fr/viz-bin/VizieR} database was
searched for additional period determinations. The Optical Gravitational Lensing
Experiment (OGLE)-II catalogue provided pulsation periods for a few more
objects, in particular among the short-period, low-amplitude variables. In cases
where two periods were available, the one from \citet{Vanh07} was adopted. For
the Plaut stars, the periods and K-band magnitudes were taken from \citet{Utt07}
and \citet{Utt08}. The mean K-band magnitudes and adopted periods are listed in
columns~4 and 5, respectively, of Table~\ref{tab_params}.

\begin{table*}
\centering
\begin{minipage}{150mm}
\caption{Main parameters of the sample stars. Meaning of the columns: object
name; Galactic longitude $l$; Galactic latitude $b$; mean K-band magnitude;
pulsation period; adopted effective temperature; luminosity derived by
integrating the SED; luminosity corrected for bulge depth scatter using the
period -- K-magnitude relations of \citet[][ see Sect.~\ref{lumi}]{Ita04};
current pulsation mass; adopted $\log g$ of the model atmosphere used in the
analysis (see Sect.~\ref{logg}); LAV: large-amplitude variable. Unknown values
are replaced by '...'. The luminosities of the comparison stars BMB~78, BMB~289,
and BD$-$012971 are calculated from their temperatures and surface gravities,
assuming 1\,$M_{\sun}$.}
\label{tab_params}
\begin{tabular}{lrrrrrrrrrl}
\hline
Object name\footnote{$^{\rm a}$: Stellar parameters from \citet{CS06}. $^{\rm b}$: Stellar parameters from \citet{RO05}.} & $l$    & $b$     & $\left<K\right>$ & Period & $T_{\rm eff}$ & $L_{\star}/L_{\sun}$ & $L_{\star}/L_{\sun}$ & $M/M_{\sun}$ & $\log g$ & Remark\\
            & [deg]  & [deg]   & (mag)            & (days) & (K)          &  uncorr.\         & corr.\            &             &  cgs     &       \\
\hline
BMB 78$^{\rm a}$     &  1.15  & $-3.78$ & 7.53 & \dots & 3600 &  654 & \dots & \dots& 0.80 & \\
BMB 289$^{\rm a}$    &  1.22  & $-4.02$ & 6.17 & \dots & 3375 & 1268 & \dots & \dots& 0.40 & \\
BMB 289$^{\rm b}$    &        &         & 6.17 & \dots & 3200 &  814 & \dots & \dots& 0.50 & \\
BD$-$012971$^{\rm b}$&$-11.5$ &  50.8   & 4.29 & \dots & 3600 & 1304 & \dots & \dots& 0.50 & \\
J174117.5-282957   &$-0.13$ &   1.04  & 7.10 & \dots  & 2800 & 5609 & \dots &\dots & $-0.28$ & \\
J174123.6-282723   &$-0.08$ &   1.04  & 8.32 & \dots  & 3200 & 1638 & \dots &\dots &   0.13  & \\
J174127.3-282851   &$-0.09$ &   1.02  & 7.26 & 438.95 & 2600 & 4732 & \dots & 1.85 & $-0.26$ & LAV \\
J174127.9-282816   &$-0.08$ &   1.02  & 7.14 & \dots  & 3000 & 4921 & \dots &\dots & $-0.20$ & \\
J174128.5-282733   &$-0.07$ &   1.02  & 7.48 & 309.46 & 3000 & 3652 &  4804 & 1.37 & $-0.20$ & \\
J174139.5-282428   &$-0.01$ &   1.02  & 7.04 & \dots  & 2800 & 5314 & \dots &\dots & $-0.28$ & \\
J174140.0-282521   &$-0.02$ &   1.01  & 7.26 & \dots  & 2900 & 4813 & \dots &\dots & $-0.20$ & \\
J174155.3-281638   &  0.14  &   1.04  & 6.82 & \dots  & 2700 & 6324 & \dots &\dots & $-0.28$ & \\
J174157.6-282237   &  0.06  &   0.98  & 7.58 & \dots  & 3300 & 3342 & \dots &\dots & $-0.13$ & \\
J174203.7-281729   &  0.14  &   1.00  & 7.55 & 392.12 & 2800 & 4541 & \dots & 1.38 & $-0.23$ & LAV \\
J174206.9-281832   &  0.13  &   0.98  & 7.26 & 481.61 & 2800 & 5152 & \dots & 1.29 & $-0.28$ & LAV \\
J174917.0-293502   &$-0.14$ & $-1.02$ & 8.55 & \dots  & 3200 & 1541 & \dots &\dots &   0.16  & \\
J174924.1-293522   &$-0.13$ & $-1.04$ & 8.54 & \dots  & 3700 & 1680 & \dots &\dots &   0.37  & \\
J174943.7-292154   &  0.10  & $-0.99$ & 7.94 & \dots  & 3000 & 2413 & \dots &\dots & $-0.15$ & \\
J174951.7-292108   &  0.12  & $-1.01$ & 8.26 & \dots  & 3500 & 1795 & \dots &\dots &   0.25  & \\
J175432.0-295326   &  0.18  & $-2.16$ & 7.29 & 401.74 & 2900 & 3844 & \dots & 0.94 & $-0.20$ & \\
J175456.8-294157   &  0.39  & $-2.14$ & 6.90 & 322.11 & 2900 & 5699 &  5410 & 1.82 & $-0.23$ & \\
J175459.0-294701   &  0.32  & $-2.19$ & 6.96 & 472.15 & 2600 & 6095 & \dots & 2.38 & $-0.26$ & LAV \\
J175515.4-294122   &  0.43  & $-2.19$ & 8.26 & ~76.90 & 3300 & 1652 &  2059 & 0.69 &   0.00  & \\
J175517.0-294131   &  0.43  & $-2.20$ & 7.99 & \dots  & 3500 & 2483 & \dots &\dots &   0.11  & \\
J180238.8-295954   &  0.96  & $-3.73$ & 7.86 & 119.00 & 3200 & 2005 &  3843 & 0.92 & $-0.21$ & \\
J180248.9-295430   &  1.05  & $-3.72$ & 8.13 & ~75.69 & 3300 & 1759 &  2116 & 0.68 &   0.00  & \\
J180249.5-295853   &  0.99  & $-3.76$ & 8.65 & ~65.18 & 3400 & 1214 &  1848 & 0.73 &   0.12  & \\
J180259.6-300254   &  0.95  & $-3.82$ & 7.46 & 228.87 & 3000 & 2971 &  3374 & 1.22 & $-0.14$ & \\
J180301.6-300001   &  1.00  & $-3.81$ & 8.37 & ~33.51 & 3400 & 1330 &   606 & 0.45 &   0.38  & \\
J180304.8-295258   &  1.10  & $-3.76$ & 8.53 & ~35.52 & 3400 & 1147 &   683 & 0.48 &   0.38  & \\
J180305.3-295515   &  1.07  & $-3.78$ & 7.15 & 207.02 & 3100 & 4023 &  2944 & 1.02 & $-0.10$ & \\
J180305.4-295527   &  1.07  & $-3.78$ & 8.00 & 113.94 & 3200 & 2015 &  4184 & 1.15 & $-0.14$ & \\
J180308.2-295747   &  1.04  & $-3.81$ & 7.92 & 113.04 & 2800 & 1919 &  3613 & 2.13 & $-0.04$ & \\
J180308.7-295220   &  1.12  & $-3.77$ & 7.94 & ~98.00 & 3100 & 1782 &  2842 & 1.04 & $-0.07$ & \\
J180311.5-295747   &  1.05  & $-3.82$ & 7.39 & 292.04 & 2900 & 3036 &  4557 & 1.64 & $-0.20$ & \\
J180313.9-295621   &  1.07  & $-3.82$ & 8.04 & ~75.00 & 3400 & 1801 &  2073 & 0.61 &   0.05  & \\
J180316.1-295538   &  1.09  & $-3.82$ & 8.31 & ~62.69 & 3300 & 1328 &  1519 & 0.65 &   0.13  & \\
J180323.9-295410   &  1.12  & $-3.83$ & 8.15 & ~84.84 & 3200 & 1547 &  2409 & 0.82 & $-0.05$ & \\
J180328.4-295545   &  1.11  & $-3.86$ & 7.40 & 119.09 & 3200 & 3005 &  3850 & 1.15 & $-0.11$ & \\
J180333.3-295911   &  1.06  & $-3.90$ & 8.44 & 203.95 & 3300 & 1247 &  2970 & 0.78 & $-0.11$ & \\
J180334.1-295958   &  1.05  & $-3.91$ & 7.06 & 119.09 & 3100 & 3986 &  3696 & 1.06 & $-0.18$ & \\
Plaut 3-45         &$-0.94$ & $-7.22$ & 6.65 & 271.02 & 3000 & 5308 &  3938 & 1.25 & $-0.19$ & LAV \\
Plaut 3-100        &$-3.85$ & $-8.92$ & 6.37 & 298.70 & 3000 & 8288 &  5432 & 1.68 & $-0.21$ & LAV \\
Plaut 3-315        &  0.04  & $-7.65$ & 6.54 & 326.80 & 3100 & 6351 &  5514 & 1.30 & $-0.27$ & LAV \\
Plaut 3-626        &$-2.63$ &$-10.10$ & 7.19 & 298.48 & 3400 & 3835 &  5344 & 0.89 & $-0.26$ & LAV, Tc-rich \\
Plaut 3-794        &  0.92  & $-9.03$ & 6.04 & 303.54 & 3000 &11278 &  5584 & 1.71 & $-0.21$ & LAV \\
Plaut 3-942        &$-2.41$ &$-11.47$ & 6.55 & 338.00 & 3500 & 6753 &  6235 & 0.83 & $-0.31$ & LAV, Tc-rich \\
Plaut 3-1147       &$-1.84$ &$-12.31$ & 5.77 & 395.63 & 3000 &13506 &  7577 & 1.84 & $-0.31$ & LAV, Tc-rich \\
Plaut 3-1347       &  0.35  &$-12.60$ & 5.99 & 426.60 & 3300 &10730 &  8143 & 1.16 & $-0.38$ & LAV, Tc-rich \\
\hline
\end{tabular}
\end{minipage}
\end{table*}

\subsubsection{Temperature}\label{temperature}

The most important stellar parameter for our spectroscopic analysis is the
effective temperature. To this end, low-resolution flux-calibrated optical
spectra were obtained by one of us (PW) with the Double-Beam Spectrograph (DBS)
mounted to the Australian National University 2.3\,m telescope at Siding Spring
Observatory (Australia). More information on the instrument can be found in
\citet{Rod88}. Exposure times ranged between 200 and 1200\,s. The precise
wavelength range of the optical spectra varies, but the part between
666 -- 1000\,nm is always covered. This range includes prominent TiO and VO
bands that are very sensitive to the effective temperature of the star
\citep[see e.g.][]{Rei05}. Furthermore, the general slope of the spectra can
constrain the star's temperature.

For the Spitzer sample stars identified as non-variable by
\citet[][Sect.~\ref{puls_period}]{Vanh07}, low-resolution spectra that were
obtained several years before the CRIRES run were used. These older spectra
have a dispersion of 0.41\,nm/pixel. The sample stars identified as variable
were observed with the DBS between 16 and 18 days after the CRIRES run in which
the Spitzer sub-sample was observed. These spectra have a somewhat higher
resolution of 0.187\,nm/pixel. We may safely assume that the temperature derived
from the optical spectra is representative for the temperature at the time of
the CRIRES observations: stars with short pulsation periods and low amplitudes
do not exhibit a large temperature change over a pulsation cycle, while for the
stars with longer periods and larger amplitudes the time delay between CRIRES
and DBS observations is small compared to the pulsation period. For the Plaut
stars, the CRIRES and DBS observations were done at independent phases, hence
here the temperature determined from the low-resolution spectra may deviate
somewhat more from that at the actual time of CRIRES observations. All sample
stars were identified to be oxygen-rich from the low-resolution spectra, i.e.\
they all exhibit the characteristic TiO bands. Also the Tc-rich stars have very
prominent TiO bands, the ZrO bands are present but not dominant. In the observed
spectral range, C-stars would be identified by intense bands of the CN molecule.

Before temperature determination, the optical low-resolution spectra were
corrected for interstellar reddening. The same extinction values as in
\citet{Vanh07} were adopted, which stem from the maps of \citet{Sch99},
\citet{Sum04}, and \citet{Mar06}. These values were applied to establish the
extinction as a function of wavelength, using the relations given in
\citet{ODo94} and \citet{Car89}. The spectra were divided by this extinction
curve to obtain the de-reddened spectrum. Test runs show that the resulting
temperatures are quite insensitive to the extinction value.

The main method to derive temperatures from the optical spectra involves fitting
of series of model spectra to the observed spectra. A series of COMARCS model
atmospheres and spectra (see Sect.~\ref{model_atmos}) was calculated, with
temperatures between 2600 and 3800\,K in steps of 100\,K,
$\log g\,[{\rm cm\,s^{-2}}]= 0.0$, 1\,M$_{\sun}$, microturbulence 3\,km\,s$^{-1}$,
solar C/O ratio, and two metallicities: 1.0\,Z$_{\sun}$ and 0.3\,Z$_{\sun}$.
Extensive lists of atomic and molecular lines, in particular of TiO and VO, were
taken into account in the calculation of the spectra. In addition, we used a
grid of PHOENIX model spectra \citep{Hus13} with temperatures between 2000 and
3900\,K in steps of 100\,K, solar metallicity, and otherwise identical
parameters as the COMARCS grid. We fitted the model spectra to the observed
spectra with a modified $\chi^2$ method in the form
\[
\chi=\frac{1}{N}\sum_{i=1}^{N}\frac{\sqrt{\left(f_{{\rm obs},i}-f_{{\rm model},i}([{\rm Fe}/{\rm H}])\right)^2}}{\frac{1}{2}\left(f_{{\rm obs},i}+f_{{\rm model},i}([{\rm Fe}/{\rm H}])\right)}.
\]
The wavelength range between 700 and 930\,nm was used to find the temperature at
which the model spectra best fit an observed spectrum. The lower wavelength
limit was chosen to avoid the noise-dominated blue part of the observed
spectrum, whereas the upper limit was chosen to avoid a molecular band of ZrO
that might be present in the stars that had experienced 3DUP but not in stars
that had not. This band could potentially have a systematic impact on the
derived temperature. For the non-variable stars in the sample, the $(V-K)_{0}$
vs.\ $T_{\rm eff}$ calibration of \citet{vanB99} was used to derive another
temperature estimate. This photometric temperature estimate agrees well with
that derived from fitting the optical spectra. The final adopted temperature
listed in column~6 of Table~\ref{tab_params} was chosen based on a visual
inspection of the observed spectra and of the model fits to them. We also tried
to derive temperatures from the spectral parameters defined in
\citet[][their Eq.~17 -- 19]{Flu94}, but found that these estimates sometimes
vary considerable for a given star. The narrow bandpasses lie at the bottom of
molecular bands that are formed in the very outer layers of the star, which are
most affected by dynamic effects (pulsation, mass loss).

The determined temperature may depend on the adopted metallicity of the model
spectra. Conversely, the metallicity that we derive from the high-resolution
spectra depends on the model input temperature. An independent determination of
effective temperature and metallicity without any assumptions is very difficult
for our sample stars. In the above procedure we find relatively modest
differences between the temperatures derived using the models with
1.0\,Z$_{\sun}$ and 0.3\,Z$_{\sun}$, respectively. The impact of the assumption on
the metallicity for temperature determination seems limited and we are thus
confident that we can derive reliable abundances from the high-resolution
spectra. In any case, the CO lines are quite insensitive to the effective
temperature.

\subsubsection{Luminosity}\label{lumi}

\begin{figure}
  \centering
  \includegraphics[width=\columnwidth,bb=81 369 538 699, clip]{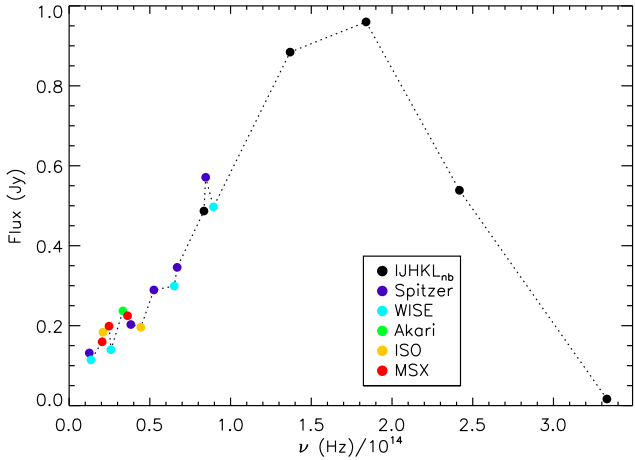}
  \caption{Typical example of a de-reddened SED of a sample star
    (J174157.6-282237). The source of the various data points is indicated in
    the legend. The luminosity of this star was found to be 3342\,$L_{\sun}$.}
  \label{SED_ex}
\end{figure}

The luminosity of each sample star was determined by integrating its spectral
energy distribution (SED). To construct the SED, we collected I-band photometry
from the DENIS catalogue \citep{Epc97}, cycle-averaged J, H, K, L$_{\rm nb}$ band
photometry from \citet{Wood98} and \citet{Vanh07}, {\it Spitzer} IRAC and
MIPS\,24 photometry \citep{Utt10,Hinz09}, as well as photometry from the WISE
\citep{Wri10}, ISOGAL \citep{Omo03}, Akari \citep{Ishi10}, and MSX
\citep{Egan03} surveys. Because for most stars the bulk of the stellar flux is
emitted between the J and L bands and because cycle-averaged fluxes in these
bands are used, we are confident to have very precise luminosities at hand for
our sample stars. The photometry was de-reddened for interstellar extinction
(see Sect.~\ref{temperature}) and converted to flux values (Jansky). The whole
SED was integrated numerically by connecting the data points by straight lines,
and an extrapolation to the origin $(\nu,f)=(0,0)$ was included. To convert this
apparent flux to a luminosity in solar units, a distance modulus of 14\fm5,
corresponding to a distance of 8.0\,kpc to the bulge, was assumed. A typical
example of a de-reddened SED is shown in Fig.~\ref{SED_ex}.

\begin{figure}
  \centering
  \includegraphics[width=\columnwidth,bb=83 369 546 699, clip]{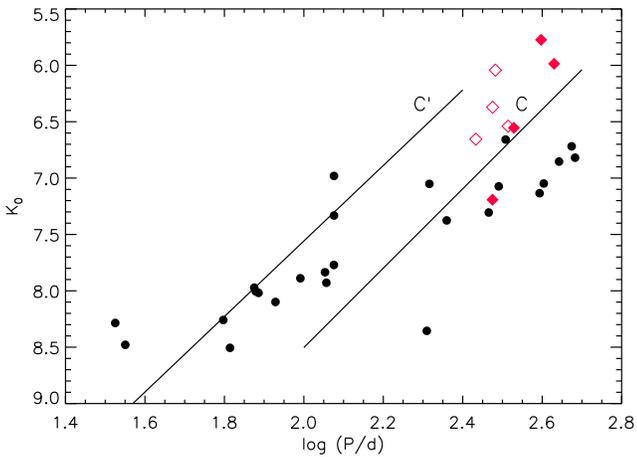}
  \caption{$\log P - K_0$ diagram of our sample stars. Black symbols are the
    Spitzer sample stars, filled red diamonds the Tc-rich Plaut stars, open red
    diamonds the Tc-poor Plaut stars. The solid lines are the relations C
    (fundamental mode pulsators) and C' (first overtone pulsators) from
    \citet{Ita04}.}
  \label{logPK}
\end{figure}

\begin{figure}
  \centering
  \includegraphics[width=\columnwidth,bb=62 366 552 700]{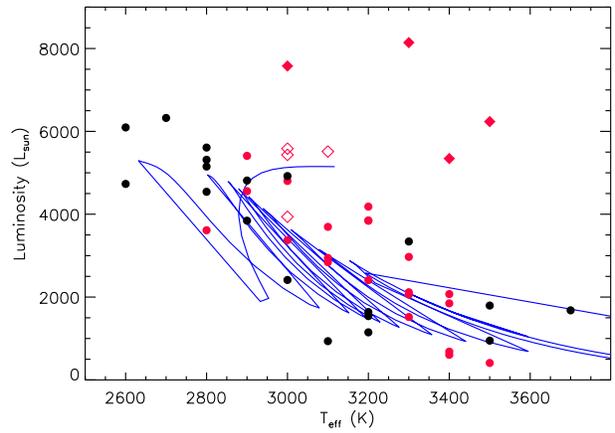}
  \caption{Luminosity vs.\ effective temperature diagram of the sample stars.
    Stars whose luminosities were not corrected for the bulge depth scatter due
    to the lack of a known pulsation period are plotted as black symbols, those
    whose luminosities were corrected are plotted as red symbols. The Plaut
    stars are plotted with the same symbols as in Fig.~\ref{logPK}. The blue
    graph is an isochrone from \citet{Bre12} with 5.0\,Gyr age, solar
    metallicity, and including dust formation on the AGB.}
  \label{HR_diagr}
\end{figure}

For stars with a measured pulsation period, we corrected these luminosities by
forcing them to fall on one of the period -- K-magnitude relations of
\citet{Ita04}, in an attempt to correct for the distance scatter within the
bulge (red points in Fig.~\ref{HR_diagr}). A difference in the distance modulus
between the Large Magellanic Cloud and the bulge of 4\fm0 was assumed. A
$\log P - K_0$ diagram (Fig.~\ref{logPK}) reveals that the long period stars
probably follow sequence C (fundamental mode or Mira-like pulsators), whereas
the shorter period stars probably follow sequence C' (first overtone or
semi-regular pulsators). We assume that the transition occurs at a period of
160\,d ($\log(P)\sim2.2$). Both the uncorrected and corrected luminosities are
listed in columns~7 and 8 of Table~\ref{tab_params}.

This correction would lead to very high luminosities (8000\,$L_{\sun}$ and more)
and subsequently unrealistically high current masses ($3-5M_{\sun}$, see next
Section) for the five Spitzer sample stars with the longest pulsation periods,
four of them being the LAVs. Because their very red $(J-K)_0$ colours (up to
2.29) and considerable mid-IR excesses suggest that their K-band magnitudes are
diminished by strong circumstellar extinction \citep[cf.\ Fig.~20 in][]{VW93} we
did not apply the distance correction to their luminosities. The long-period
Plaut stars do not show such extreme colours, which is why the correction was
applied nevertheless.

The Plaut stars preferentially lie above sequence C in Fig.~\ref{logPK}, whereas
the Spitzer stars seem to lie preferentially below the sequences. The location
of the Plaut stars can be understood as a result of selection effects that
favoured stars on the near side of the bulge. They were discovered by
photographic plate photometry \citep{Plaut} and selected for optical
spectroscopy \citep{Utt07}. The location of the Spitzer stars is more
surprising, at the moment we do not have a convincing explanation why they are
preferentially at distances farther than the Galactic centre.

Figure~\ref{HR_diagr} shows a Hertzsprung-Russell diagram with inverted
temperature axis of the sample stars. The solid blue line in that figure is an
isochrone from \citet{Bre12} with 5.0\,Gyr age, $Z=0.017$ (solar metallicity),
and including dust formation on the AGB. Stars on the AGB of this isochrone have
a current mass of $\sim1.29M_{\sun}$. The RGB tip of this isochrone is at
$\sim2500L_{\sun}$. Most of the sample stars define a clear trend of increasing
luminosity with decreasing temperature, as may be expected from stars
distributed along the AGB. Because temperature and luminosity of the stars were
derived from completely independent data, it can be excluded that a systematic
effect produces such a trend. The isochrone covers well the distribution of of
our sample stars and reproduces the trend of increasing luminosity with
decreasing temperature. The Tc-rich Plaut stars are shifted away from the
isochrone to somewhat higher temperature and/or higher luminosity. At lower
metallicity, stars on the AGB have higher temperatures at a given luminosity. As
will be shown later, the two Tc-rich stars for which we are able to determine
the metallicity are relatively metal-poor. This could explain their displacement
from the location of the other stars.

\subsubsection{Current mass and surface gravity}\label{logg}

{\em Current} pulsation masses for the variable stars with a determined period
were computed. The linear pulsation models used here have been applied to LMC
cluster variables, where they yield masses that are in good agreement with those
expected from the cluster ages determined by main-sequence turn-off fitting
\citep{Kam10}. Linear pulsation periods were computed with the code described in
\citet{LW07}, except that the low temperature opacities were generated from the
web site described in \citet{MA09}. Metal and helium mass fractions of $Z=0.016$
and $Y=0.28$ were used. A grid of AGB models with $600<L/L_{\sun}<10\,000$ and
$0.7<M/M_{\sun}<5.0$ was made and the linear fundamental and first overtone
periods calculated. Analytic fits to the models were made in the form
\[
\log\frac{M}{M_{\sun}}=0.324-1.699p+3.050r+1.007p^2+1.891r^2-2.515pr
\]
and
\[
\log\frac{M}{M_{\sun}}=-0.143-2.148p+3.271r-0.687p^2-0.894r^2+1.524pr
\]
for the fundamental and first overtone modes, respectively, where $p=\log P-2$
and $r = 0.5\left(\log L/L_{\sun} - 4\log T_{\rm eff}\right) + 5.3$. These fits to
the pulsation mass are accurate to better than 7\% for all computed models.
Stars with $L/L_{\sun}<2000$ could be RGB stars rather than AGB stars. Similar
fits were made to RGB models with luminosities $600<L/L_{\sun}<2000$ and
$0.6<M/M_{\sun}<1.0$, the difference between the AGB and RGB models being the
core mass. The core masses for the AGB models were assumed to be given by the
formula in \citet{WZ81} while RGB masses were obtained from the stellar models
of \citet{Gir00}. The RGB fits are
\[
\log\frac{M}{M_{\sun}}=0.351-1.900p+3.397r+1.303p^2+2.917r^2-3.750pr
\]
and
\[
\log\frac{M}{M_{\sun}}=-0.165-2.527p+3.682r-1.256p^2-0.952r^2+2.343pr
\]
for the fundamental and first overtone modes, respectively, with an accuracy
better than 2\%.

Using the values of $P$, $L/L_{\sun}$ (corrected with the help of period -- 
K-magnitude relations), and $T_{\rm eff}$ given in Table~\ref{tab_params},
pulsation masses were computed from the above formulae. It was assumed that
stars with $P<160$\,d are first overtone pulsators and those with $P>160$\,d are
fundamental mode pulsators (cf.\ Fig.~\ref{logPK}). The pulsation masses are
listed in column~9 of Table~\ref{tab_params}. Since for stars with
$L/L_{\sun}<2000$ the difference between RGB and AGB pulsation masses is small,
only the AGB pulsation mass was adopted. The masses of the five Spitzer sample
stars with the longest pulsation periods have to be considered the least
certain.

With these current masses estimated from pulsation theory, the surface gravity
of the sample stars was calculated using the ``classical'' relation:
\[
\log g_{\star} = \log g_{\sun} + 4\log\left(\frac{T_{\star}}{T_{\sun}}\right) -
\log\left(\frac{L_{\star}}{L_{\sun}}\right) +
\log\left(\frac{M_{\star}}{M_{\sun}}\right),
\]
adopting $T_{\sun}=5770$\,K and $\log g_{\sun}=4.44$.
A few stars turned out to have unrealistically low masses ($M<0.7M_{\sun}$),
i.e.\ masses less than the carbon-oxygen core plus a thin envelope that is
expected to be left when the star leaves the tip of the AGB. To obtain more
realistic $\log g$ values for them, a current mass of $0.7M_{\sun}$ was adopted.
For stars without a known pulsation period, a current mass of $M=0.85M_{\sun}$
was adopted as a first guess. This is a reasonable assumption for a supposedly
old population. However, for most of these stars, this lead to a very low
surface gravity, lower than for what the COMARCS atmospheric models converged
(Sect.~\ref{model_atmos}). These stars are also outliers in a $\log g$ vs.\
luminosity diagram. Thus, for them we used the most extended model atmospheres
that were available (i.e.\ lowest $\log g$). These adopted $\log g$ values are
well within the range expected from the general trend of $\log g$ vs.\
luminosity formed by the other stars. Column~10 of Table~\ref{tab_params} lists
the surface gravities $\log g$ finally adopted for the model atmospheres.

\subsection{Model atmospheres}\label{model_atmos}

Model atmospheres for our sample stars were established with the help of the
COMARCS code. COMARCS is based on a revised version of MARCS \citep{Jor92} with
spherical radiative transfer routines from \citet{Nor84} and assumes local
thermodynamic equilibrium (LTE). The opacities are taken from tables prepared
by the COMA code \citep{Ari00}. Recent COMARCS models are described in
\citet{Ari09}.

\subsection{Measuring radial velocities and abundances}\label{rv_abu}

Radial velocities (RVs) of the stars were determined from the numerous CO
$\Delta\nu=2$ lines in the K-band setting in grating order 24 by
cross-correlation with a generic synthetic model spectrum. These lines are quite
insensitive to abundances and stellar parameters, hence the choice of the model
has a negligible impact on the measured radial velocities. The typical
uncertainty on the RV is $\sim1$\,km\,s$^{-1}$, but may be better for the
hotter, less variable sample stars. The radial velocities are summarised in 
column~2 of Table~\ref{tab_results}.

\begin{table*}
\centering
\begin{minipage}{125mm}
\caption{Radial velocities and abundances measured from the CRIRES spectra.
Meaning of the columns: object name; heliocentric radial velocity; overall
[Fe/H] metallicity; C/O ratio; carbon isotopic ratio; relative abundances of Al,
Si, Ti, and Y. Abundances of stars for which no acceptable fit was achieved are 
omitted (\dots), abundances followed by a colon are uncertain (poor fit).}
\label{tab_results}
\begin{tabular}{lrrrrrrrr}
\hline
Object name\footnote{$^{\rm a}$: Abundances adopting the stellar parameters from \citet{CS06}. $^{\rm b}$: Abundances adopting the stellar parameters from \citet{RO05}.}      & RV$_{\rm helio}$ & [Fe/H] & C/O & $^{12}$C/$^{13}$C & [Al/Fe] & [Si/Fe] &  [Ti/Fe] & [Y/Fe] \\
                 & (km\,s$^{-1}$) &        &     &                 &     &     &     &     \\
\hline
BMB 78$^{\rm a}$   &  --59.7 & --0.47 & 0.46  &   9.9 & & & & \\
BMB 289$^{\rm a}$  &  --65.2 & --0.05 & 0.57  &  4.0: & & & & \\
BMB 289$^{\rm b}$  &         & --0.21 & 0.67  &  4.6: & & & & \\
BD$-$012971$^{\rm b}$&--119.6& --0.68 & 0.30  &  9.5 & $+0.40$ & $+0.21$ & $+0.60$ & $-0.07$ \\
J174117.5-282957 &   159.4 & \dots   & \dots &   9.3:&  & & &  \\
J174123.6-282723 &    66.5 & --0.07  & 0.39  &  18.2 & $+0.05$ & $-0.04$ & $+0.36$ & $-0.02$ \\
J174127.3-282851 &    58.7 & \dots   & \dots & \dots &  & & &  \\
J174127.9-282816 & --166.0 &   0.22  & 0.22  &  14.5 & $-0.03$ & $-0.10$ & $+0.17$ & $+0.11$ \\
J174128.5-282733 &    15.0 & --0.21  & 0.25  &  14.6 &  \dots  & \dots   & \dots   & \dots   \\
J174139.5-282428 &   102.3 & \dots   & \dots &  4.8: &  & & &  \\
J174140.0-282521 &  --89.7 & \dots   & \dots & 11.3: &  & & &  \\
J174155.3-281638 &   177.8 & \dots   & \dots & 11.9: &  & & &  \\
J174157.6-282237 & --101.7 & --0.18  & 0.42  &   5.5 & $+0.09$ & $+0.16$ & $+0.50$ & $+0.19$ \\
J174203.7-281729 & --234.2 & \dots   & \dots & \dots &  & & &  \\
J174206.9-281832 &    56.2 & \dots   & \dots & \dots &  & & &  \\
J174917.0-293502 &    18.1 &  +0.01  & 0.50  &  19.4 & $+0.18$ & $-0.07$ & $+0.20$ &   0.00  \\
J174924.1-293522 &     2.2 & --0.10  & 0.23  &  18.3 &  & & &  \\
J174943.7-292154 &   --9.3 & --0.16  & 0.44  &  22.2 & $-0.05$ & $-0.02$ & $+0.23$ & $-0.18$ \\
J174951.7-292108 &  --72.4 & --0.06  & 0.51  &  17.2 &  & & &  \\
J175432.0-295326 &    15.5 & \dots   & \dots & 17.7: &  & & &  \\
J175456.8-294157 &   107.1 & \dots   & \dots & 17.4: &  & & &  \\
J175459.0-294701 &   198.6 & \dots   & \dots &  9.8: &  & & &  \\
J175515.4-294122 &   171.0 &  +0.01  & 0.44  &  16.5 &  & & &  \\
J175517.0-294131 &  --55.0 & --0.08  & 0.34  &  16.9 &  & & &  \\
J180238.8-295954 &   103.2 & --0.17  & 0.45  &  16.0 & $+0.25$ & $+0.23$ & $+0.43$ & $+0.13$ \\
J180248.9-295430 &   143.3 & --0.15  & 0.48  &  20.7 & $+0.21$ & $+0.01$ & $+0.41$ & $+0.19$ \\
J180249.5-295853 &  --42.3 & --0.30  & 0.29  &  12.8 &  & & &  \\
J180259.6-300254 &  --40.3 & --0.33  & 0.26  &  22.3 &  & & &  \\
J180301.6-300001 &    58.4 & --0.05  & 0.46  &  19.3 &  & & &  \\
J180304.8-295258 &    33.7 &   0.07  & 0.33  &   9.9 &  & & &  \\
J180305.3-295515 &  --21.4 & --0.27  & 0.30  &   9.8 & $+0.17$ & $+0.18$ & $+0.44$ & $+0.15$ \\
J180305.4-295527 &    27.5 & --0.18  & 0.26  &  15.5 & $+0.16$ & $+0.18$ & $+0.39$ & $+0.02$ \\
J180308.2-295747 &    14.9 & \dots   & \dots & 11.0: & \dots   & \dots   & \dots   & \dots   \\
J180308.7-295220 &    13.7 & --0.12  & 0.51  &  27.9 & $+0.23$ & $+0.11$ & $+0.38$ & $+0.02$ \\
J180311.5-295747 &    44.8 & \dots   & \dots & \dots & \dots   & \dots   & \dots   & \dots   \\
J180313.9-295621 &  --96.4 & --0.26  & 0.29  &  11.0 & $+0.15$ & $+0.04$ & $+0.23$ & $-0.11$ \\
J180316.1-295538 &    62.5 & --0.15  & 0.47  &  18.4 &  & & &  \\
J180323.9-295410 &  --42.4 & --0.21  & 0.37  &  18.8 & $+0.11$ & $+0.01$ & $+0.21$ & $-0.07$ \\
J180328.4-295545 & --113.7 & --0.17  & 0.32  &   7.5 &  & & &  \\
J180333.3-295911 &     6.6 & --0.56  & 0.39  &   9.7 &  & & &  \\
J180334.1-295958 &    24.0 & --0.11  & 0.47  &  16.1 & $+0.31$ & $+0.27$ & $+0.35$ & $+0.09$ \\
Plaut 3-45       &    16.3 & \dots   & \dots & \dots &         &         &         &         \\
Plaut 3-100      &  --22.9 & \dots   & \dots & \dots &         &         &         &         \\
Plaut 3-315      &  --61.4 & \dots   & \dots & \dots &         &         &         &         \\
Plaut 3-626      &  --67.2 & $-0.85$ & 0.71  & \dots &         &         &         &         \\
Plaut 3-794      &  --50.1 & \dots   & \dots & \dots &         &         &         &         \\
Plaut 3-942      &  --80.1 & $-0.56$ & 0.75  & 19.6: &         &         &         &         \\
Plaut 3-1147     &   --2.3 & \dots   & \dots & \dots &         &         &         &         \\
Plaut 3-1347     &    63.0 & \dots   & \dots & \dots &         &         &         &         \\
\hline
\end{tabular}
\end{minipage}
\end{table*}

COMARCS model atmospheres adopting the main stellar parameters
($T_{\rm eff},\log g$) derived in Sect.~\ref{stellar_params} were calculated for
each star, spanning a grid of various general abundance combinations. In a first
try, we established a three-dimensional grid in [Fe/H], C/O, and [$\alpha$/Fe],
where the $\alpha$-elements are O, Ne, Mg, Si, S, Ar, Ca, and Ti. [Fe/H] varied
between $-1.5$ and $+0.5$ in steps of 0.5, for C/O the values 0.15, 0.30, 0.60,
and 0.90 were adopted, and [$\alpha$/Fe] varied between $-0.20$ and $+0.40$ in
steps of 0.20. The solar abundances listed in \citet{Caf08} were adopted as
reference scale. The solar metallicity on that scale is $Z_{\sun} = 0.0156$ and
the solar C/O ratio is 0.55. A generic microturbulence of 2.5\,km\,s$^{-1}$, a
value typical for evolved giants, was adopted for the model calculation
\citep{Cun07}\footnote{This value is slightly lower than the microturbulence
used in the grid used for temperature determination, but the difference has
virtually no impact on the results.}. Based on these model atmospheres,
synthetic spectra were generated for the wavelength range of the observed H-band
spectra. Important line lists that were taken into account were the atomic line
list of \citet{Ryde10}, the CN as well as the H$_2$O line lists from
\citet{Jor97}\footnote{This H$_2$O line list was chosen due to its
completeness.}, the OH list from the HITRAN 2004 compilation \citep{Rot05}, and
the CO list from \citet{GC94}. For both the CN and CO line lists the wavelengths
of some transitions were corrected with the help of line identifications of
\citet{Hin95}. The synthetic spectra with an original resolving power of
$R=300\,000$ were convolved with a Gaussian profile to that of the observed
spectra at $R=50\,000$. We developed an IDL routine that interpolated between
these grid spectra and fitted the interpolated model spectrum to the observed
one of each star. This interpolation was done for every point of the wavelength
vector. The standard IDL routine {\tt amoeba.pro} was employed to find the
interpolated abundance combination that achieved the best fit to the observed
spectrum, by minimising a $\chi^2$ criterion.

\begin{figure}
  \centering
  \includegraphics[width=\columnwidth,bb=83 369 538 700]{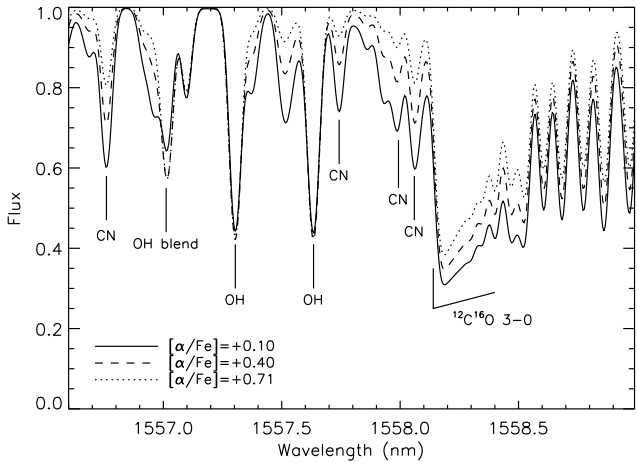}
  \caption{Synthetic spectra based on COMARCS atmospheres with
    $T_{\rm eff}=3500$\,K, $\log g= +0.25$, $[{\rm Fe}/{\rm H}]=+0.16$, and
    $[\alpha/{\rm Fe}]$ ratios as indicated in the legend. The model atmospheres
    and synthetic spectra were consistently calculated with these abundances
    while keeping the carbon abundance constant.}
  \label{OHl}
\end{figure}

The initial aim was to derive the abundance of oxygen, the most abundant
$\alpha$-element, from lines of the OH molecule that are prominently present in
the H-band. However, it turned out that the OH lines present in the spectra are
not sensitive enough to use them as reliable abundance indicator of O. In a
consistent spectral synthesis (i.e.\ the same chemical abundances are adopted
for the model atmosphere and spectral synthesis calculation), the strength of
the OH lines is almost unchanged, even in a wide range of O abundances. Such a
low sensitivity of molecular line strength to abundance is known to occur in
cool stellar atmospheres and can be explained by the feedback of the
abundance of key elements, most importantly O, on the atmospheric structure
due to their impact on molecular opacities. The atmospheric structure may change
just in such a way as to leave OH line strengths unchanged. This point is
illustrated in Fig.~\ref{OHl}, which shows a zoom-in on synthetic spectra around
three OH lines just blue-ward of the $^{12}$CO 3-0 band head. The CO and CN
lines become weaker with increasing oxygen abundance because the chemical
equilibrium is changed to the disfavour of these molecules, whereas the OH lines
remain unchanged with O abundance. This conclusion from our models was also
checked and confirmed with models from the PHOENIX code \citep{Hus13}. Hence,
from the OH lines available in our spectra the O abundance cannot be derived
with a sufficient precision in cool, M-type giants, and all previous such
determinations that used an inconsistent calculation have to be taken with
caution. Only for hotter stars such as Arcturus ($T_{\rm eff}\gtrsim4000$\,K) the
OH lines are sensitive to the O abundance.

Because it is not possible to constrain the $\alpha$-abundance from the H-band
spectra, the original 3D grid of model atmospheres had to be reduced to a 2D
grid. Here we make the crucial {\it assumption} that [$\alpha$/Fe] is a function
of [Fe/H] as follows: $[\alpha/{\rm Fe}]=+0.40$ at [Fe/H]=$-1.5$, $-1.0$, and
$-0.5$, followed by a linear decrease of 0.4\,dex/dex up to [Fe/H]=+0.5. This
run closely approximates that found from dwarf and sub-giant
\citep[][see their Fig.~25]{Bens13} as well as giant bulge stars
\citep[e.g.][]{Alv10}.

In addition to fitting the two detector chips from the H-band setting, chip~2
from the additional K-band setting grating order 26 was included when available.
This chip was chosen because it contains a number of metal lines that are not
too strong, but only a relatively small number of CN lines. The metal lines
should provide for a better constraint on the metallicity of the stars, while
the CN lines are also sensitive to the C/O ratio. The change in [Fe/H] and C/O
implied by including this additional piece of K-band spectrum is relatively
small. For the 13 science targets with reliable abundance determination (i.e.\
excluding J174128.5-282733, J180308.2-295747, and J180311.5-295747), the
average change is 0.00\,dex, with a standard deviation of 0.07\,dex.

A map of $\chi^2$ in the ([Fe/H], C/O) plane shows that the $\chi^2$ values form
a relatively flat valley that runs diagonally through that plane: A similarly
good fit as at the minimum is achieved at a lower (higher) metallicity, if at
the same time the C/O ratio is increased (decreased). By including the
additional chip from the K-band, the walls of the ``valley'' become steeper
along the [Fe/H] axis, which means that the metallicity is better constrained
for the stars with these additional observations.

A first test of the analysis methods was done with synthetic observations, i.e.\
synthetic spectra that were created with model atmospheres that have abundance
combinations that are between the model grid. Gaussian noise was added to the
spectra to simulate an ${\rm S}/{\rm N}\sim40$. Also, a slightly altered line
list was used for the synthetic observations to simulate absorption lines
missing in the models. It turned out that the fitting routine tended to converge
to a metallicity that was slightly ($\sim0.10$\,dex) lower than the input
metallicity, and a C/O ratio that was slightly higher than the input C/O. The
results improved considerably when wavelength points were neglected from the
fitting procedure whose flux deviated too much ($\sim0.10$ of continuum flux)
from the observed flux at a given iteration step. Hence, this clipping technique
was also adopted for the analysis of the real spectra.

The results of the fitting procedure were checked by visually inspecting a
number of metal lines that were found to be sensitive to [Fe/H], but not to C/O
(i.e., free of CN and CO blends). This included the following lines: Fe 1549.46,
Fe 1553.60, Fe 1553.85, Fe 1557.10, and Ni 1561.00\,nm. If these lines were not
well reproduced by the automatic fitting procedure, the fitting was repeated
manually to reproduce the strength of these lines. Subsequently, the fitting
procedure was run again to measure the C/O ratio of the star with the
metallicity fixed to the value derived manually.

\begin{figure*}[ht!]
  \centering
  \includegraphics[width=17cm,bb=66 378 566 574]{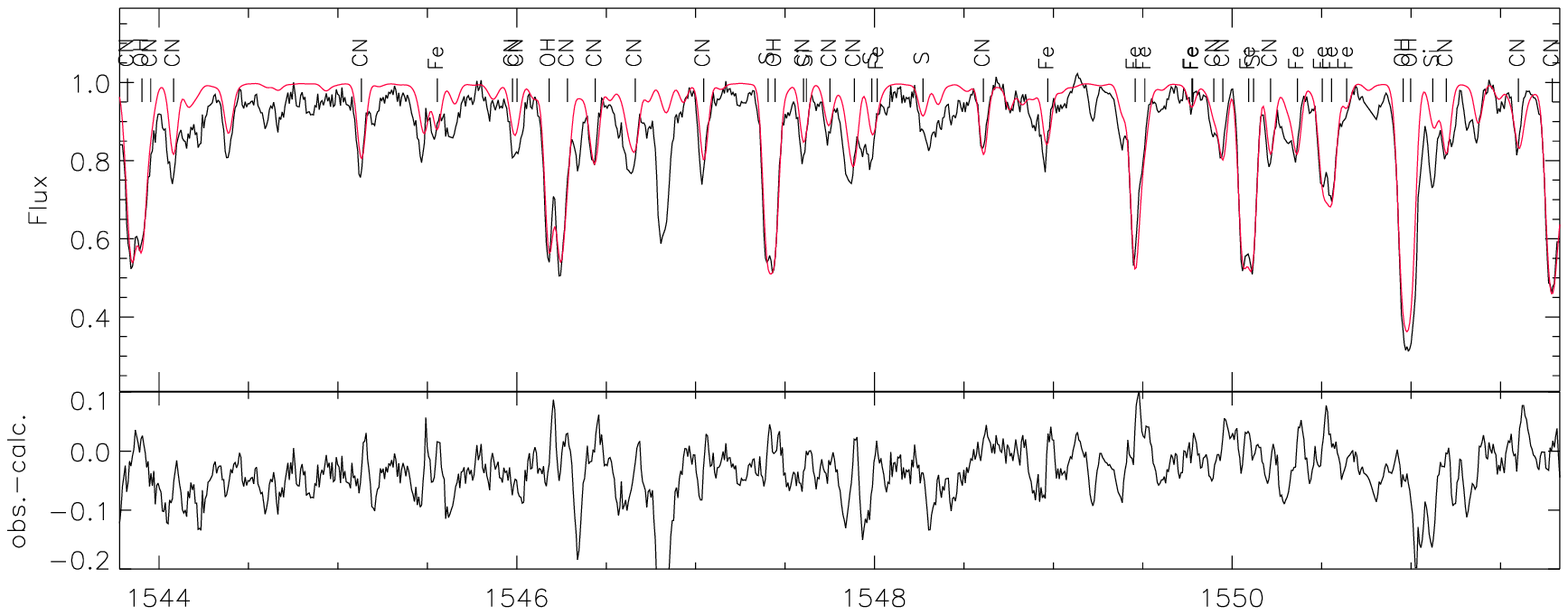}
  \includegraphics[width=17cm,bb=66 378 566 578]{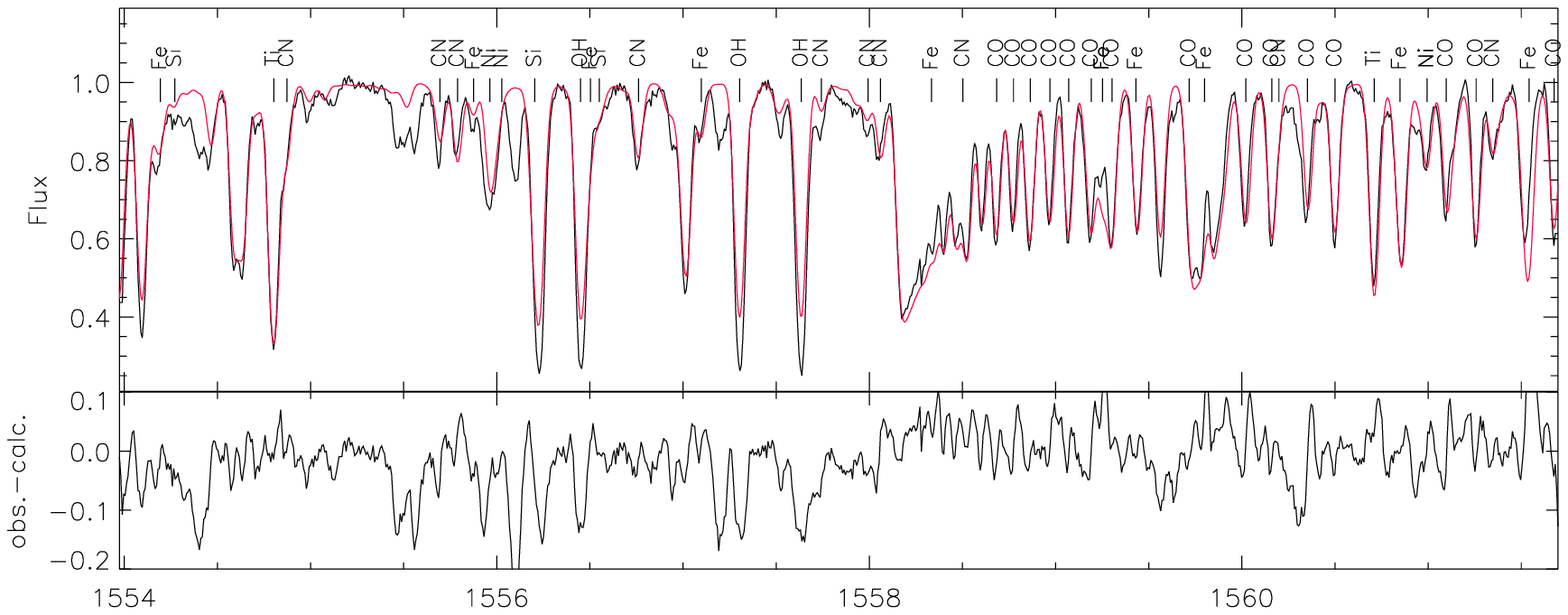}
  \includegraphics[width=17cm,bb=66 366 566 578]{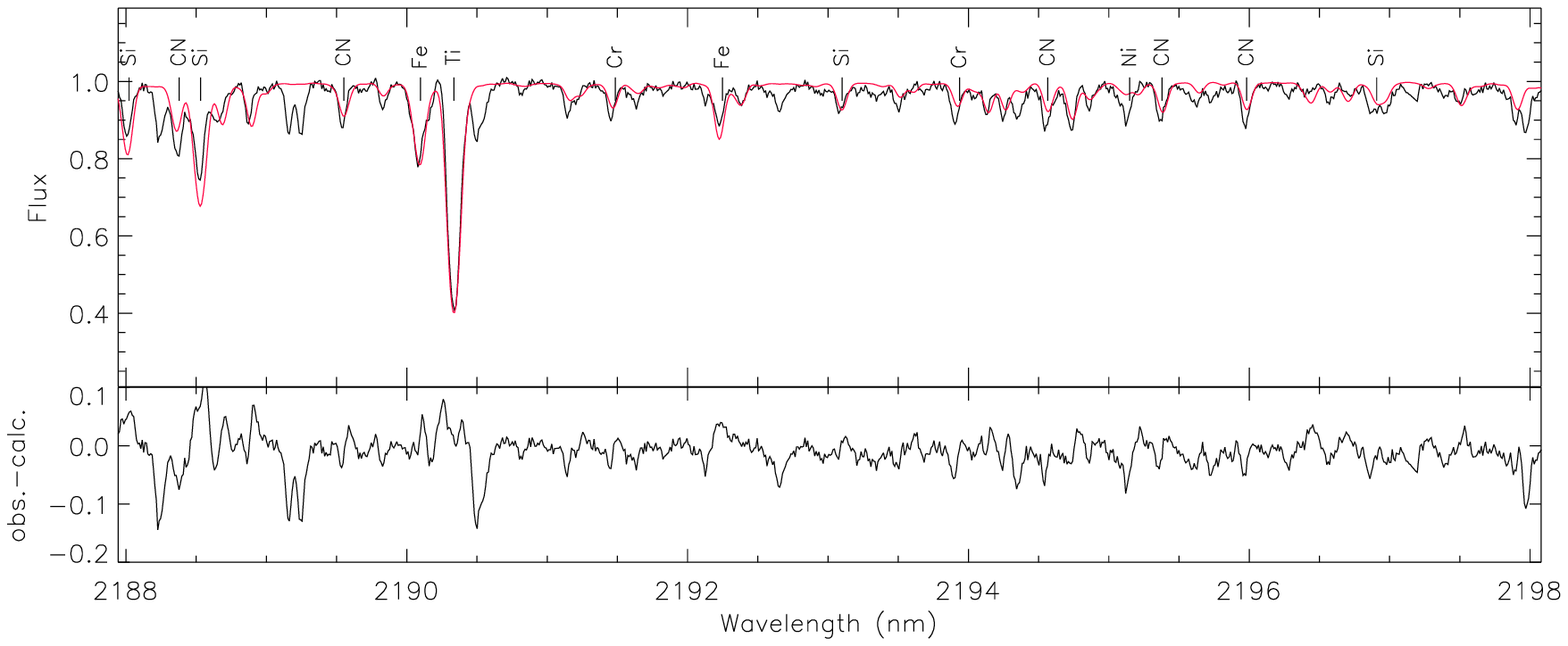}
  \caption{Typical example of a spectral fit achieved in this work, for star
    J180313.9-295621. The upper, middle, and lower panel show chip~2 and 3 of
    the H-band setting (order 36), and chip~2 from the K-band setting with order
    26, respectively. The upper sub-panel in each panel shows in black the
    observed spectrum and in red the best-fitting interpolated synthetic
    spectrum. The lower sub-panel in each panel shows the residuals
    (observed -- calculated flux) of the fit. This star was found to have
    $T_{\rm eff}=3400$\,K, $\log g=+0.05$, $[{\rm Fe}/{\rm H}]=-0.26$, and
    C/O=0.29.}
  \label{example_fit}
\end{figure*}

A typical example of a spectral fit is shown in Fig.~\ref{example_fit}. No
dynamic effects such as line splitting, emission features, etc., are seen in
the warmer ($T_{\rm eff}\geqslant3100$\,K) stars, even in the variable ones. Most
of the large residuals (observed minus calculated flux) do not vary in
wavelength from star to star; they are most probably caused by missing line
data, cf.\ \citet{Ryde09} for some unidentified features.

We may thus assume that the ability to model the spectra of our sample stars is
mainly limited by the (in)completeness of available line lists, except for the
coolest stars ($T_{\rm eff}\lesssim3000$\,K), for which the hydrostatic models
are not realistic and water starts to play an important role. At this low a
temperature, the metal lines lose their power to constrain the metallicity and
the results for these cool stars are not reliable. Thus, abundances for stars
cooler than $T_{\rm eff}=3000$\,K are not reported in Table~\ref{tab_results}.

Most notably, it is not possible to model the spectra of the four LAVs in the
Spitzer sample, even though they were observed close to minimum light when their
spectra are supposed to be least complex. A model spectrum using only the line
list of water by \citet{Rot10} was calculated, based on one of the cool models
($T_{\rm eff}=2800$\,K) to compare it to observed cool star spectra. Many of the
strong lines in the observed spectra have a counterpart in the synthetic water
spectrum, suggesting that water is indeed a strong contributor to the spectrum
of these stars. This comparison also shows that the match in wavelength between
observed spectrum and water line list is good at this high a resolution.

The abundance determination was however successful for the two warmest Plaut
stars, even though they are large amplitude variables.

We note that our analysis represents a significant effort. The computation of a
single opacity table with one abundance combination takes 19 hours on a fast,
modern desktop PC. The construction of the model atmospheres requires
significant human interaction because the models do not easily converge at low
$\log g$. A more detailed analysis is beyond the resources available to us
currently.

\subsection{The carbon isotopic ratio and metal abundances}

To determine abundances of metals as well as the carbon isotopic ratio, we
established for each star a tailored model atmosphere adopting the stellar
parameters derived in Sect.~\ref{stellar_params} and assuming the general
abundance pattern ([Fe/H], C/O) as found from the H-band spectra (and chip~2
of order 26, where available).

\begin{figure}
  \centering
  \includegraphics[width=\columnwidth,bb=84 370 536 699]{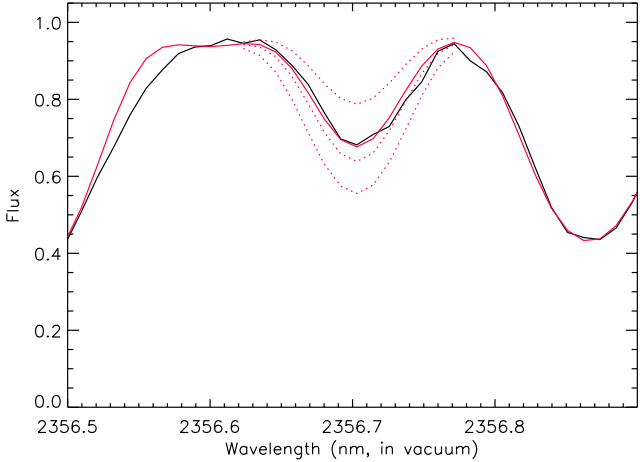}
  \caption{Fit to the $^{13}$CO \mbox{2-0} R76 line of the star
    J174123.6-282723. The solid red line shows the synthetic spectrum
    interpolated to the best-fit $^{12}$C/$^{13}$C=18.2. The dotted lines show
    the synthetic spectra calculated with $^{12}$C/$^{13}$C=50, 12, and 3.5
    (from top to bottom).}
  \label{ex12C13C}
\end{figure}

The carbon isotopic ratio was derived from lines of $^{13}$CO located in chip~1
of grating order 24. An inspection of the optical depth of line cores as
a function of gas temperature reveals that the low-J (R19, R22, R24, R26)
$^{13}$CO \mbox{2-0} lines are formed relatively far out in the atmosphere where
dynamic effects may cause significant deviations of the model atmosphere from
the real atmosphere. The analysis was thus based on the high-J \mbox{2-0} lines
(R76, R78, R80, R81). These lines are also more sensitive to changes in
$^{12}$C/$^{13}$C than the low-excitation lines. Spectra with $^{12}$C/$^{13}$C
ratios of 3.5 (equilibrium value of the carbon isotopic ratio in the CN
cycling), 7, 12, 20, 30, 50, and 90 (solar value) were synthesised based on the
tailored model atmospheres. The isotopic ratio of the stars cooler than 3000\,K,
for which no general abundance pattern could be determined, was measured with a
generic abundance combination of [Fe/H]=0.0, $[\alpha/{\rm Fe}]=+0.20$, and
C/O=0.30. Their isotopic ratios are more uncertain and are marked by a colon in
Table~\ref{tab_results}. Again, an interpolation and fitting routine was
employed to determine the value of isotopic ratio that best fits the observed
spectrum in the mentioned lines. Figure~\ref{ex12C13C} shows a typical example
fit to the $^{13}$CO \mbox{2-0} R76 line of the star J174123.6-282723. Note the
good fit also to the $^{12}$CO \mbox{4-2} R64 line at the right hand side of the
figure, which gives additional confidence in the C/O ratio that was derived for
this star (0.39), though the CO $\Delta\nu=2$ lines in the K-band are generally
less sensitive to C/O than the CO $\Delta\nu=3$ lines in the H-band.

We also aimed at deriving abundances of individual metals using equivalent width
(EW) analysis. Weak enough ($\log({\rm EW_{obs}}/\lambda)<-4.5$) but unblended
lines of Al, Si, Ti, and Y were found in the spectra in grating orders 26 and
27. Y is a heavy element produced in the slow neutron-capture (s-) process, an
increase in its abundance can indicate recent or ongoing 3DUP activity in an AGB
star. Astrophysical gf values of the lines were established with the Arcturus
spectrum \citep{Hin95}, adopting the abundances for Al, Si, and Ti reported in
\citet{RAP11}, and a scaled solar abundance for Y. Other atomic parameters were
taken from the VALD database \citep{Kup00}. The resulting line list is reported
in Table~\ref{linelist}. Synthetic spectra with varying abundances of the
individual elements were calculated from the tailored model atmospheres, from
which the EWs of the respective lines were measured. The abundance in each star
was then calculated by interpolating in the $\log({\rm EW_{synth}}/\lambda)$ vs.\
$\log\epsilon({\rm X})$ plane to the observed reduced EW
($\log({\rm EW_{obs}}/\lambda)$).

\begin{table}
\centering
\caption{Line list used in the determination of the metal abundances. Meaning of
  the columns: chemical symbol of the element; wavenumber of the transition;
  oscillator strength gf; excitation energy of the lower level in
  $10^{4}{\rm cm}^{-1}$.}
\label{linelist}
\begin{tabular}{lccc}
\hline
Element & $\sigma$   & gf       & $\xi$                 \\
        & (cm$^{-1}$) &          & ($10^4{\rm cm}^{-1}$)  \\
\hline
Al\,I   & 4713.8730  & 3.291e--1 & 4.13194 \\
Al\,I   & 4713.8810  & 3.291e--1 & 4.13194 \\
Si\,I   & 4779.4609  & 3.104e+0  & 5.42569 \\
Ti\,I   & 4589.4991  & 1.256e--1 & 1.41066 \\
Y\,I    & 4702.2860  & 1.000e--1 & 1.15337 \\
\hline
\end{tabular}
\end{table}

\subsection{Abundance error estimate}\label{err_est}

To estimate the formal errors on the measured abundances, uncertainties in four
quantities were taken into account: temperature, surface gravity, continuum
placement, and microturbulence.

An estimate of the temperature uncertainty may be derived from the scatter that
most stars exhibit around the general trend of decreasing temperature with
increasing luminosity in the HRD (Fig.~\ref{HR_diagr}). If the LAVs and the
obviously deviating Plaut stars are excluded, the remaining sample stars follow
a linear least squares fit in $T_{\rm eff}$ vs.\ uncorrected luminosity with a
standard deviation of 181\,K. For the stars with known pulsation period the
scatter reduces from 175\,K before correction to 119\,K after correction. Note
that the relation between temperature and luminosity may have an intrinsic
scatter and curvature (as indicated by the isochrone in Fig.~\ref{HR_diagr}),
and that the temperature is not pulsation-cycle averaged. We may thus assume
that the uncertainty of the temperature determination is not larger than this
scatter. An uncertainty of $\pm100$\,K was adopted.

To estimate the uncertainty on the surface gravity ($\log g$), we adopted the
temperature uncertainty, an uncertainty on the luminosity of 10\%, and on the
pulsation period of 1\%. These uncertainties were applied to the formulae given
in Sect.~\ref{logg}. From this, a typical uncertainty of $\Delta\log g=\pm0.12$
was derived.

The continuum placement was changed by $\pm1$\% and the microturbulence $\xi$ by
$\pm1$\,km\,s$^{-1}$. The latter seemed a reasonable uncertainty from the CO
lines in the K-band, which are quite sensitive to $\xi$. To determine the
abundance uncertainty, the model atmospheres and synthetic spectra were
recalculated, varying the parameters one at a time by the given uncertainties,
and the spectral fitting was repeated for a few representative stars. The total
uncertainty on the abundances was calculated by combining the individual sources
of error in quadrature.

The general abundance pattern ([Fe/H], C/O) turned out to be quite robust. The
metallicity is most sensitive to temperature and $\xi$, whereas the C/O ratio is
most sensitive to changes in $\log g$ and $\xi$ (the CO lines in the H-band are
quite sensitive to $\log g$). Changes in [Fe/H] due to temperature uncertainty
vary between 0.01 and as much as 0.21\,dex, but usually they are
$\sim0.06$\,dex. Typical combined uncertainties on [Fe/H] and C/O are 0.10\,dex
and 0.08, but C/O is more robust at high values ($\pm0.05$ at C/O=0.50). The C/O
ratio was found to change by about 0.1 if $\log g$ is varied by $\pm0.3$\,dex.
This means that changing the stellar mass, and hence $\log g$, within any
reasonable limits does not significantly change our results and conclusions on
the C/O ratio of the stars.

The parameter having the largest impact on the $^{12}$C/$^{13}$C ratio is the C/O
ratio. Changing it by 0.10 can change the isotopic ratio by as much as
$\sim30$\%. The errors are correlated: An overestimate (underestimate) of C/O
will also lead to an overestimate (underestimate) of $^{12}$C/$^{13}$C, because
with initially stronger (weaker) CO lines it takes less (more) $^{13}$C to
get the same line strength. The surface gravity and continuum placement have
some influence on the $^{12}$C/$^{13}$C ratio, but is limited to $\sim10$\% and
$\sim7$\%, respectively. We estimate a combined uncertainty of $\pm\sim33$\% on
the $^{12}$C/$^{13}$C ratio, comparable to literature values.

Typical uncertainties for the elemental abundances are $\pm0.11$ for [Al/Fe],
$\pm0.25$ for [Si/Fe], $\pm0.13$ for [Ti/Fe], and $\pm0.16$ for [Y/Fe].

\subsection{Comparison stars}\label{comp_stars}

To test our tools for abundance determination, we applied them to a few
comparison stars. The goal of this exercise was not to check literature results,
but rather to compare abundances derived with our tools to those from other
studies and to identify possible systematic differences.

The most important comparison object in studies of cool stars is of course
Arcturus. A grid of model atmospheres spanning the same range of [Fe/H] and C/O
values as for the science targets was calculated, adopting the stellar
parameters determined by \citet{Ryde10}, which are: $T_{\rm eff}=4280$\,K,
$\log g=1.7$, and $\xi_{\rm micro}=1.7$\,km\,s$^{-1}$. Synthetic spectra were
generated and fitted to the observed, high-resolution spectrum provided by
\citet{Hin95} in the same spectral range in which the science targets were
observed with CRIRES. The fitting routine converged at a metallicity of
$[{\rm Fe}/{\rm H}]=-0.50$, which agrees very well with the value of
$[{\rm Fe}/{\rm H}]=-0.53\pm0.05$ derived by \citet{Ryde10}. Our C/O ratio of
0.32 is somewhat higher than 0.24 as found by these authors, whereas
\citet{RO05} determined C/O=0.28 in Arcturus. Since Arcturus is a K-type giant
whose atmospheric structure is less influenced by molecular opacity than that of
M-type giants, we also performed the 3D fit introduced in Sect.~\ref{rv_abu}.
With this approach the following abundance pattern for Arcturus was found:
$[{\rm Fe}/{\rm H}]=-0.51$, C/O=0.37, and $[\alpha/{\rm Fe}]=+0.31$. The
$\alpha$-abundance ratio is in excellent agreement with the value of +0.33 found
by \citet{Ryde10} from an average of the abundance of O, Si, S, and Ti. We also
modelled the observed CRIRES spectra of the comparison stars BMB~78, BMB~289
(both located in the Galactic bulge), and BD$-$012971 (a solar neighbourhood
star). The stellar parameters of these stars were not determined as for the
science targets, rather the parameters provided by \citet{CS06} and
\citet{RO05}, respectively, were adopted (see Table~\ref{tab_params}). The
results of our abundance determination as well as the literature values are
summarised in Table~\ref{tab_comps}.

\begin{table*}
\centering
\begin{minipage}{118mm}
\caption{Comparison of abundances with literature results.}
\label{tab_comps}
\begin{tabular}{lccccccc}
\hline
Name                 & [Fe/H]   & [Fe/H] & C/O       & C/O   & $^{12}$C/$^{13}$C & $^{12}$C/$^{13}$C & Ref.\footnote{References: CS06 = \citet{CS06}, RO05 = \citet{RO05},
  RS15 = \citet{RS15}.} \\
                     & this work & lit.\ & this work & lit.\ & this work       & lit.\           & \\
\hline
BMB 78      & $-0.47$ & $-0.03$ & 0.46    & 0.20  & 9.9  & \dots       & CS06 \\
BMB 289     & $-0.05$ & $-0.05$ & 0.57    & 0.28  & 4.0: & \dots       & CS06 \\
BMB 289     & $-0.21$ & $-0.15$ & 0.67    & 0.12  & 4.6: & $5.0\pm1.2$ & RO05 \\
BD$-$012971 & $-0.68$ & $-0.24$ & 0.30    & 0.15  & 9.5  & $7.9\pm1.8$ & RO05 \\
BD$-$012971 & $-0.68$ & $-0.78$ & 0.30    & \dots & 9.5  & \dots       & RS15 \\
\hline
\end{tabular}
\end{minipage}
\end{table*}

The differences to the literature values vary from star to star. Our [Fe/H] is
lower than the literature [Fe/H] value by more than 0.4\,dex for BMB~78 and
BD$-$012971, whereas for BMB~289 the agreement in the metallicity is excellent
\citep[the fit to the observed spectrum of that star is somewhat better with the
parameters of][]{RO05}. However, the metallicity of BD$-$012971 was recently
also determined independently by \citet{RS15} using different data to be
$[{\rm Fe}/{\rm H}]=-0.78$, in good agreement with our result
($[{\rm Fe}/{\rm H}]=-0.68$).

The C/O ratio found in our modelling is higher than the literature value by
large factors in some cases. To identify the reason for this difference, the
original spectra of the comparison stars obtained by \citet{RO05} and
\citet{CS06} were kindly made available to us by these authors. A comparison
showed that the spectra of BMB~78 and BD$-$012971 are indeed very similar to
ours, i.e.\ the difference in the abundances can not be attributed to the
different observational material. Some variation can be seen for the spectra
of BMB~289. This star is identified by \citet{Sos13} to be a semi-regular
variable with a primary pulsation period of 162.53\,d and an I-band amplitude
of 0\fm132. Most likely, the difference seen in the spectra is due to dynamic
effects in the atmosphere of that star. The fit to the spectrum of BMB~289 is of
lower quality than for the other stars. Also, because of its late spectral type
\citep[M9,][]{BMB}, this star could be considerably cooler than assumed in the
literature. For these reasons, BMB~289 appears to be not a good comparison star
and we put less weight on it.

The C/O ratios reported by \citet{RO05} are very low. Using the abundance
pattern given in the literature, our models do not give a proper fit of the
observed spectra. Note that $T_{\rm eff}$ and $\log g$ as given in the literature
were adopted. The reason for the differences found in the C/O ratios must lie in
the different analysis methods. Because of the insensitivity of the OH lines
available here to the O abundance in M-type giants (Sect.~\ref{rv_abu}),
we assumed a priori a trend of the O abundance with metallicity that was derived
from hotter stars. This trend might not be applicable to our sample. However, we
believe that most of the difference in the results stems from inconsistencies in
the O abundance used for model construction and for spectral synthesis,
which may produce different C/O ratios. We should add that the measured C/O
ratios in our sample are expected to be reliable at least in a relative sense,
so that we are able to draw conclusions on the occurrence of internal C
enrichment by 3DUP.

Table~\ref{tab_comps} also collects the $^{12}$C/$^{13}$C ratios reported in the
literature \citep{RO05} and measured here. The ratio is quite uncertain for
BMB~289, but for BD$-$012971 the agreement is very good.

In conclusion, the agreement with literature results on the individual star
level is mixed. The excellent agreement that was found for the well-studied star
Arcturus however gives confidence that the abundance patterns that we derived
for our science targets are reliable. As shown in Sect.~\ref{sect_mh}, the good
agreement in mean metallicity on the level of late-type star samples in the
inner bulge regions gives us further confidence in our results.

\section{Results and discussion}\label{res_disc}

In most of the following subsections, the Plaut stars need to be discussed
separately from the Spitzer stars because they stand out from the sample in
terms of location in the bulge, but also in metallicity and C/O ratio.
Comparisons with other bulge samples are based only on the Spitzer sample
because of closer spatial overlap. We ask the reader to pay attention to which
subsample is discussed in which place.

\subsection{Current masses}\label{sec_mass}

The current masses may tell us something about the age of the sample stars and
the star formation history in the Galactic bulge. Due to stellar mass loss, the
{\em initial} masses must have been higher than the current masses. Some of the
masses listed in Table~\ref{tab_params} are significantly larger than
$1.0M_{\sun}$, implying ages considerably lower than 10\,Gyrs. This is in
agreement with what was found previously from studies of AGB stars in the bulge
\citep{vLoon03,GB05,Utt07}: luminosities, pulsation periods, and dredge-up
behaviour suggest that there are stars of masses $1.5-2.0M_{\sun}$ and
intermediate age ($\lesssim5$\,Gyrs) present in the bulge. Although not obvious
in purely photometric studies of the turn-off \citep{Zoc03,Cla08,Cla11}, recent
investigations of microlensed dwarf and sub-giant stars with known metallicities
by \citet{Bens13} indicate that a younger population is probably present in the
bulge. The masses derived for our sample stars are in line with those results.
An extended star formation history is required to explain the more massive AGB
stars in the bulge.

\begin{figure}
  \centering
  \includegraphics[width=\columnwidth,bb=68 370 548 700]{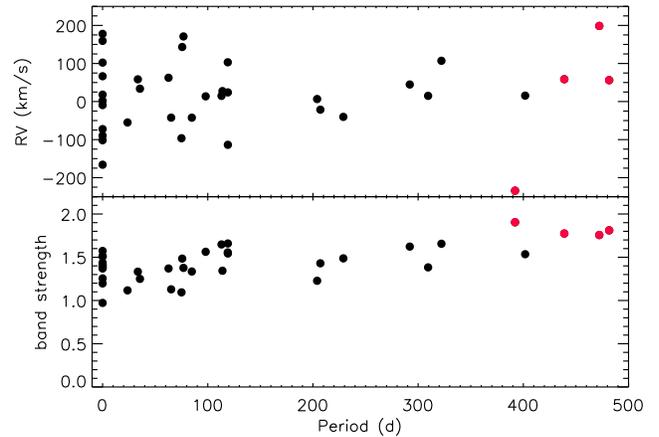}
  \caption{Radial velocity and band strength index as a function of pulsation
    period of the Spitzer sample stars. The LAVs are shown as red symbols. The
    extreme velocities as well as the highest band strengths are found among the
    stars with the longest periods. Stars for which no period is known are set
    to $P=0$.}
  \label{RVP}
\end{figure}

The four LAVs among the Spitzer sample seem to form a distinct group. They have
the following independent properties that distinguish them from the other stars:
their spectra, with very strong TiO and VO bands, make them the coolest stars
observed; their luminosities, if corrected for the depth scatter, would be the
highest of the stars in the sample; they have the longest observed pulsation
periods, from 392 to 472 days; and they have both the lowest and highest radial
velocities of any of the stars observed, suggesting that their velocity
dispersion is higher than that of the other stars. The probability to find the
extreme RV values among any sub-sample of four stars within our Spitzer sample
of 37 stars is 0.9\%. The upper panel of Fig.~\ref{RVP} shows the distribution
of the Spitzer sample stars in the RV vs.\ period diagram. The lower panel of
that figure shows a band strength index (pseudo-continuum: $911 - 921$\,nm,
band: $923 - 938$\,nm) as a function of period. This band strength index shows
a very strong linear correlation with the determined temperature. The four LAVs
are plotted as red symbols. The four LAVs not only have the highest and lowest
radial velocity, they also have the strongest molecular bands, indicating that
they form a distinct group among the Spitzer sample.

The high luminosities of these stars are consistent with their high derived
masses in that high-mass stars evolve further up the AGB before they eject all
their envelopes. The very low temperatures, especially if the stars really are
massive, suggest that these stars are of high metallicity since the giant branch
becomes cooler with increasing metallicity, but hotter with increasing mass.
Unfortunately, their near-IR spectra are too complex to derive any abundances
from them.

\citet{WB83} observed a similar group of long period variables in the bulge and
they also deduced masses of $3 - 5 M_{\sun}$. They concluded that these objects
had metallicities about $2.5Z_{\sun}$ and suggested that they could be young
metal-rich stars ejected from star-forming regions in the Galactic plane near
the centre of the Galaxy. Star clusters recently formed near the Galactic centre
are now well known \citep[e.g.][]{Wang06}. The $3 - 5 M_{\sun}$ stars could also
be ejected from the central star cluster where young stars are known to exist
\citep{GEG10}. The high RV dispersion obtained for the present sample of four
cool, high-luminosity, long-period variables lends support to this suggestion.


\subsection{Metallicity}\label{sect_mh}

\begin{figure}
  \centering
  \includegraphics[width=\columnwidth,bb=82 362 546 700, clip]{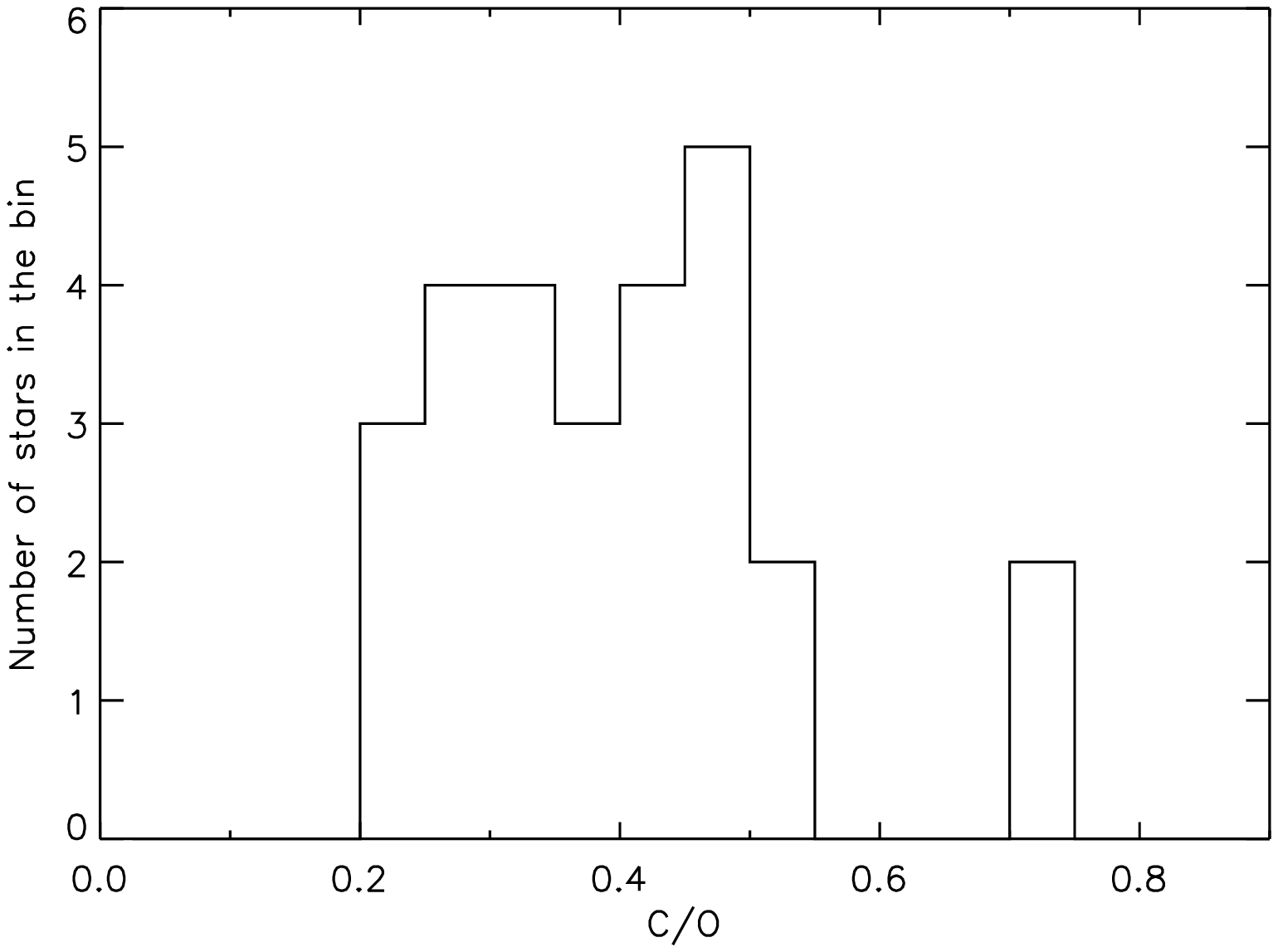}
  \includegraphics[width=\columnwidth,bb=82 405 546 700, clip]{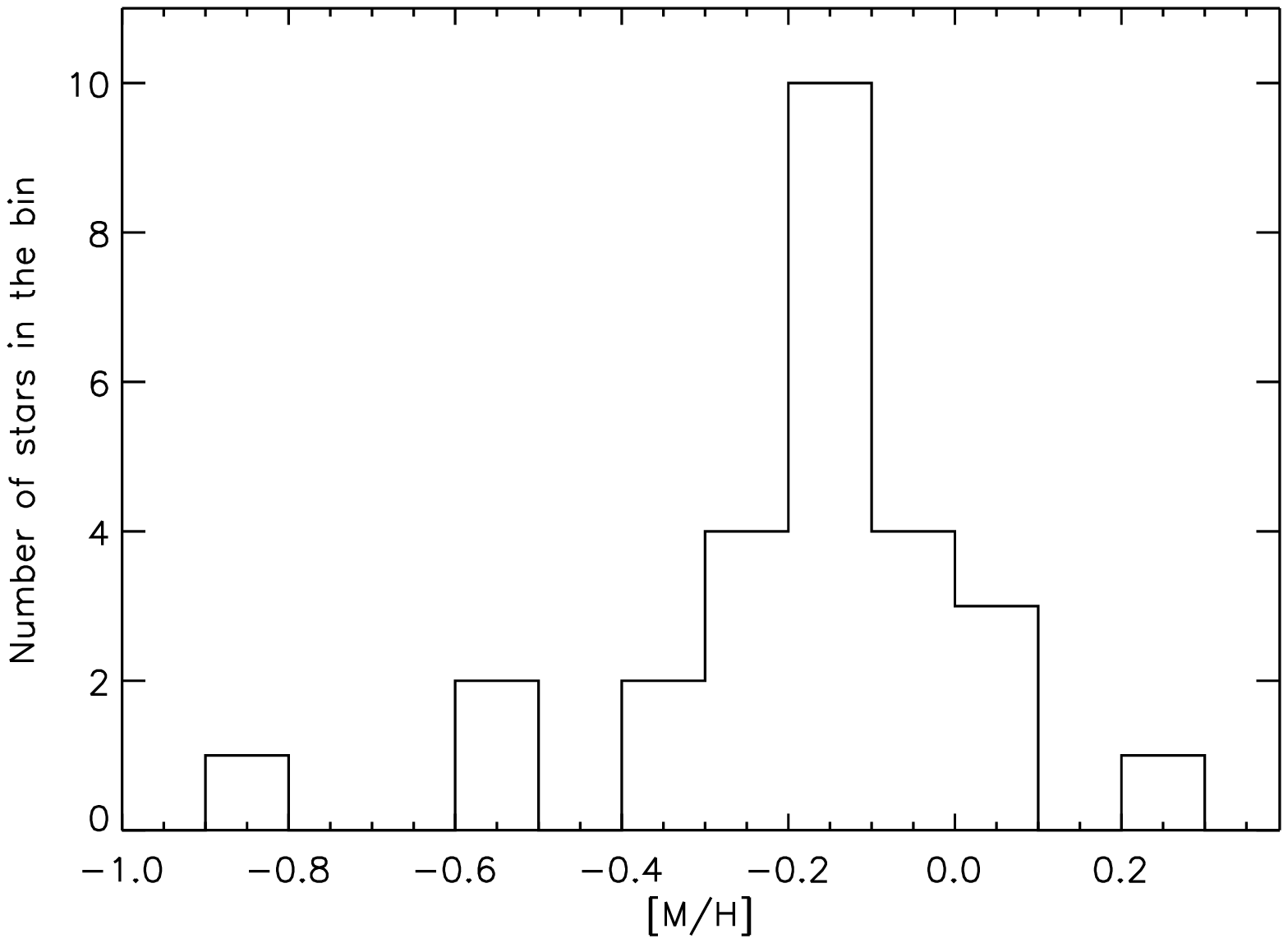}
  \includegraphics[width=\columnwidth,bb=82 369 546 699, clip]{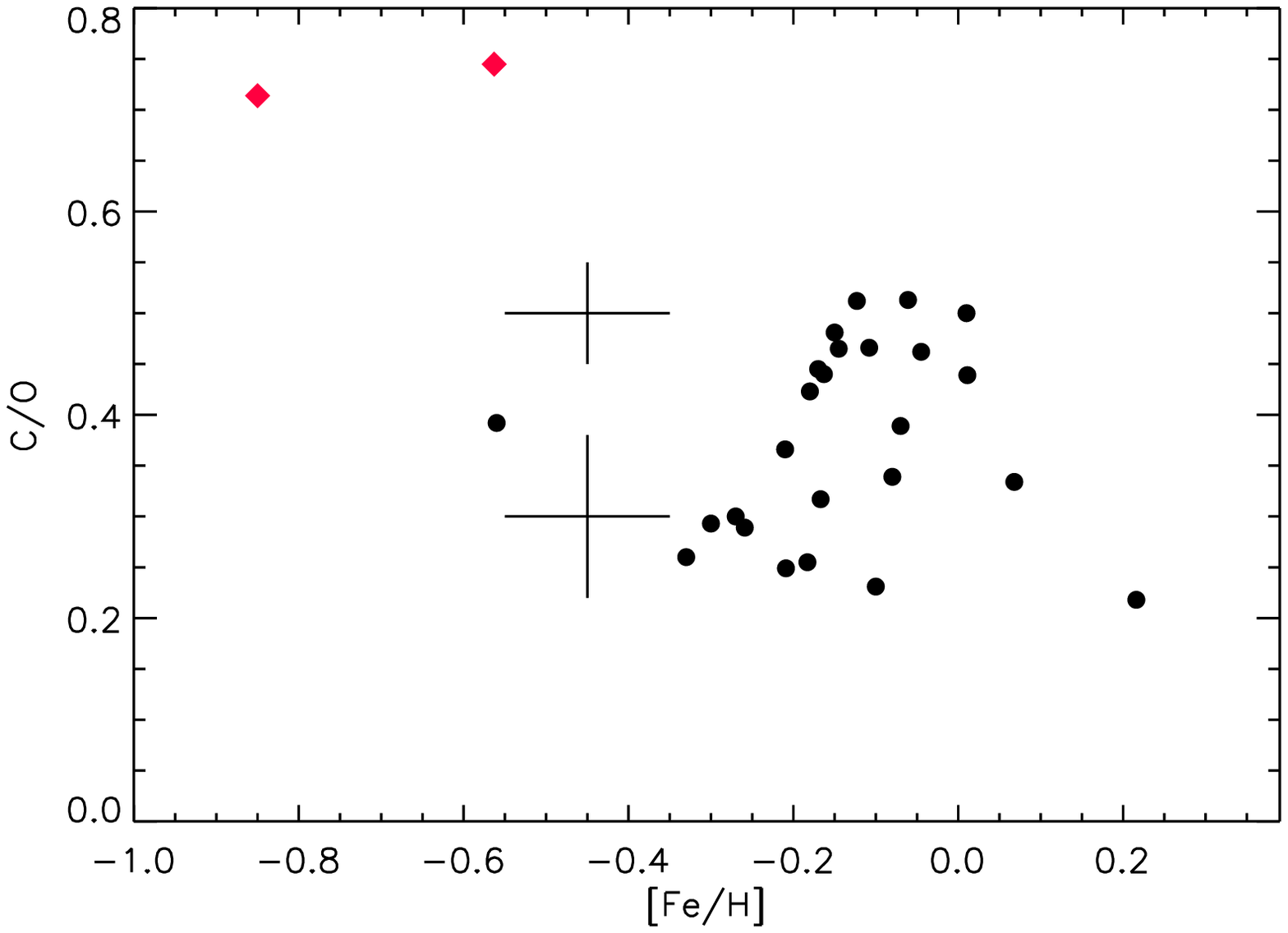}
  \caption{{\em Top panel:} Histogram of the C/O ratios. {\em Middle panel:}
    Histogram of the metallicity of all sample stars with reliable metallicity
    determination. {\em Bottom panel:} C/O ratio as a function of metallicity.
    The Plaut stars are shown as red symbols, also two typical error bars are
    shown.}
  \label{FeH_CO}
\end{figure}

The metallicity distribution function (MDF) of the sample stars is shown in the
middle panel of Fig.~\ref{FeH_CO}. The Plaut stars are included in this diagram.
The MDF is symmetric and shows a clear peak between $[{\rm M}/{\rm H}]=-0.10$
and $-0.20$.

The median metallicity of the Spitzer sample is $[{\rm M}/{\rm H}]=-0.15$, the
mean is $\left<[{\rm M}/{\rm H}]\right>=-0.14\pm0.03$, with a standard deviation
of $0.15\pm0.02$\,dex. Only few stars have super-solar metallicity. The Spitzer
sample appears to be very homogeneous. The result for the Spitzer sample may be
compared to some results from the literature: \citet{RO05} find
$\left<[{\rm Fe}/{\rm H}]\right>=-0.19\pm0.02$ from the analysis of
high-resolution ($R=25\,000$) H-band spectra of 14 M-type giants in Baade's
Window at $(l,b)=(+1\degr,-3.9\degr)$, in excellent agreement with our result.
The $1\sigma$ dispersion in their sample is $0.080\pm0.015$\,dex, somewhat
narrower than in ours. \citet{CS06} analysed high-resolution ($R=50\,000$) H-
and K-band spectra of seven bulge K- and M-type giants and derive
$\left<[{\rm Fe}/{\rm H}]\right>=-0.15$. Despite some disagreement with those
works on individual objects (Sect.~\ref{comp_stars}), the mean metallicities of
the samples agree. \citet{Rich07} observed 17 M giants in a field $1\degr$ south
of the Galactic centre and measure
$\left<[{\rm Fe}/{\rm H}]\right>=-0.22\pm0.03$ with a $1\sigma$ dispersion of
$0.140\pm0.024$\,dex. Finally, \citet{Rich12} observed a total of 30 M giants in
two fields $1.75\degr$ and $2.65\degr$ south of the Galactic centre and find in
these fields $\left<[{\rm Fe}/{\rm H}]\right>=-0.16\pm0.03$ and
$\sigma=0.12\pm0.02$, and $\left<[{\rm Fe}/{\rm H}]\right>=-0.21\pm0.02$ and
$\sigma=0.09\pm0.016$, respectively. All these results for (non-variable) K- and
M-type giants agree very nicely with our result for the Spitzer sample.

\begin{figure}
  \centering
  \includegraphics[width=\columnwidth,bb=76 369 537 698, clip]{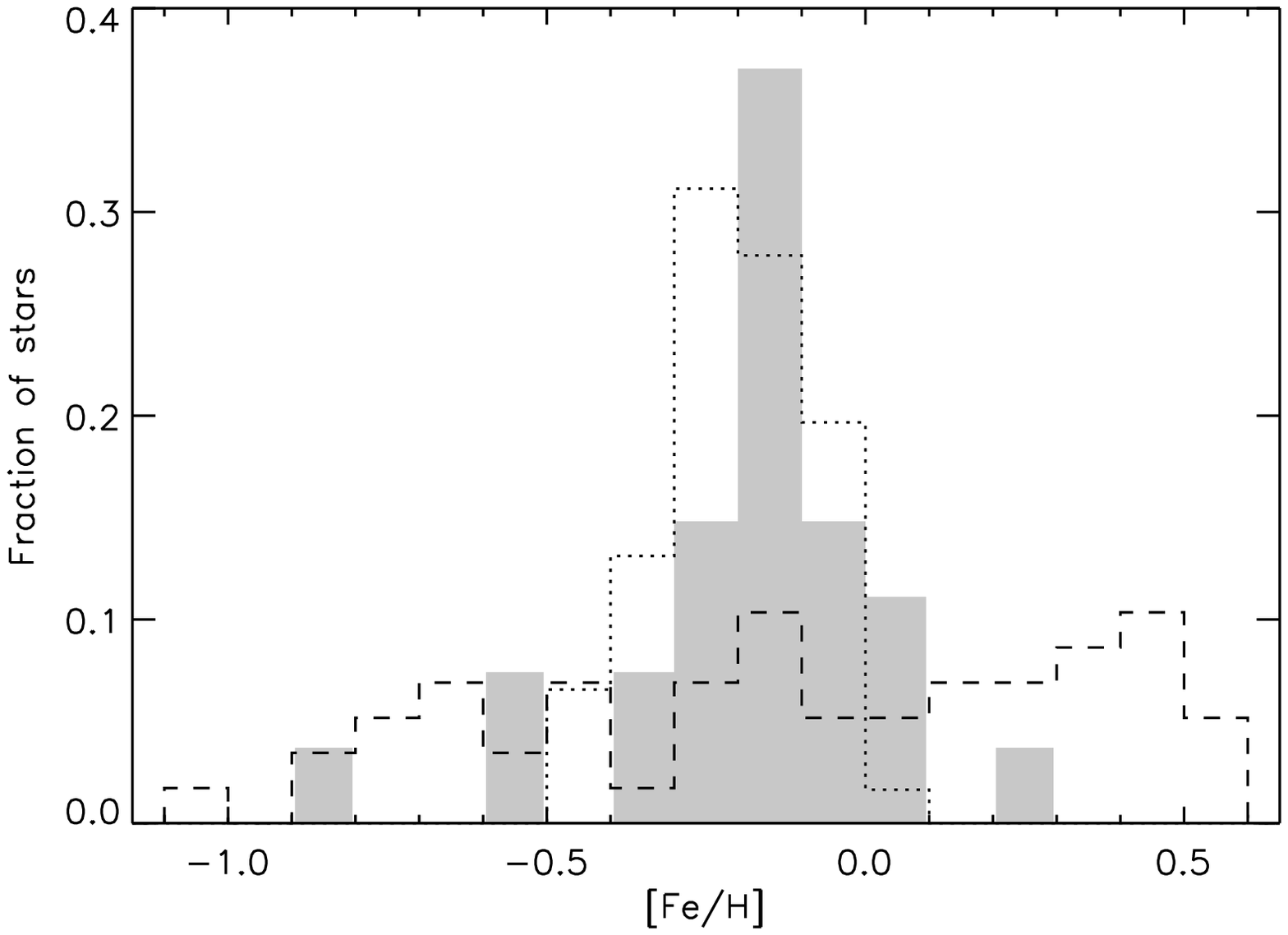}
  \caption{Comparison of MDFs: this work (shaded histogram); combined MDF from
    \citet{RO05}, \citet{Rich07}, and \citet[][dotted line]{Rich12}; and
    \citet[][dashed line]{Bens13}.}
  \label{MDF_comp}
\end{figure}

A lack of metal-rich stars among M-type giants in the bulge was noted in
previous works \citep[e.g.][and references therein]{Utt12,Rich12}. This is
surprising because only relatively metal-rich stars would be expected to become
cool enough to evolve to M-type giants. Figure~\ref{MDF_comp} shows a comparison
between MDFs of in total 61 bulge M-type giants \citep{RO05,Rich07,Rich12} and
AGB stars (27 stars, this work), and the MDF of bulge dwarf and subgiant stars
\citep[58 stars,][]{Bens13}. While the MDFs of M-giants and AGB stars are quite
narrow and peak between $[{\rm Fe}/{\rm H}]=-0.3$ and $-0.1$, the MDF of dwarfs
is much broader with less clear peaks. The last also contains more metal-rich
stars than do the giant samples. A two-sided Kolmogorov-Smirnov test reveals
that the probability that our Spitzer and the dwarf star sample are drawn from
the same underlying population is only 0.01. For the 61 M giants and the dwarf
sample, the mutual probability is even less than $10^{-5}$.

It has been suggested that very high mass-loss rates at high metallicities could
remove those stars from the canonical paths of stellar evolution
\citep{Coh08,CC93}, hence they do not reach the coolest and most advanced
phases. The low number of super-solar metallicity stars in our sample is in
agreement with this conjecture.

On the other hand, \citet{Bue13} finds that bulge PNe descend mostly from
super-solar metallicity stars. This result may be biased towards
high-metallicity objects because only PNe with relatively high surface
brightness can be studied at the distance of the bulge. Furthermore, metal-rich
stars may be more likely to form PNe than metal-poor ones, e.g.\ due to enhanced
mass loss and/or a higher binary fraction.

The two Plaut stars for which a metallicity could be determined have a lower
metallicity than the remaining sample (one Spitzer sample star has the same
metallicity as one of the Plaut stars). Though not a strong confirmation, this
finding is consistent with the conclusion of \citet{Utt12} that AGB stars in the
outer bulge ($b\approx-10\degr$) descend mainly from the metal-poor bulge
population, not from the metal-rich one. This conclusion was however solely
based on the kinematical properties (radial velocity dispersion) of AGB stars
and of the metal-poor and metal-rich population of RGB stars in a field centred
on $(l,b)=(0\degr,-10\degr)$, not on actual metallicity determinations.
\citet{Utt12} found the metal-poor population in that field to have a mean
metallicity $[{\rm M}/{\rm H}]=-0.57\pm0.03$ with a dispersion of
$0.28\pm0.03$\,dex. The two Plaut stars analysed here fall well into this
distribution, which suggests that they actually descend from this metal-poor
bulge population. Note however, that non-negligible biases could influence this
finding: Of the 671 long-period and semi-regular variables found by
\citet{Plaut} in the Palomar-Groningen field no. 3, 27 were selected for optical
spectroscopy by \citet{Utt07}, of which only eight have been observed with
CRIRES and analysed here, four of them being Tc-rich.

A negative metallicity gradient has previously been found along the minor axis
of the Galactic bulge \citep{Zoc08}. Our complete sample is in agreement with
such a gradient because the metal-poor Plaut stars are far from the plane. Even
when only the Spitzer sample is considered, we find a slight negative
metallicity gradient, but this is not very significant.

\subsection{C/O and carbon isotopic ratios -- Which AGB stars undergo 3DUP?}
\label{sect_co}

All sample stars are oxygen-rich, i.e. there is no carbon star in the
sample. Also among those sample stars whose abundances could not be measured,
none has an enhanced strength of the CO band head, except for the Plaut stars
with Tc. The distribution of C/O ratios in the sample is shown in the upper
panel of Fig.~\ref{FeH_CO}. Our sample shows an even distribution of C/O ratios
between 0.22 and 0.51, again except for the Plaut stars, which have a
considerably higher C/O ratio. This was expected because these two stars were
found to show lines of Tc, an indicator of recent or ongoing 3DUP. A comparison
between the spectra of the Plaut stars with those of similar stars from the
Spitzer sample shows that the former indeed have much more prominent CO and CN
lines. Hence, the two Plaut stars are a good benchmark of which C/O ratio we may
expect to find in stars that underwent 3DUP and to check if any of the Spitzer
sample stars underwent 3DUP.

The mixed chemistry observed in many Galactic bulge PNe \citep{Guz11} may be
understood if those objects descend from stars such as the Tc-rich Plaut stars.
They have an enhanced C/O ratio ($\gtrsim0.7$), which would ease the formation
of PAHs in their outflows once they become post-AGB objects or PNe. Our finding
of a relatively high C/O ratio in two of the Plaut stars
(${\rm C}/{\rm O}\sim0.7$) does not exclude the existence of an UV-irradiated
torus around the central stars of Galactic bulge PNe suggested by \citet{Guz11}.
Nevertheless, our observations suggest that the assumption of low C/O ratios in
PNe precursors towards the Galactic bulge is not easily justified.

The solar C/O ratio in the compilation of \citet{Caf08} used here is 0.55, hence
all Spitzer sample stars have an at least slightly sub-solar C/O ratio. The mean
C/O ratio of the Spitzer stars is 0.38, with standard deviation of 0.10. Careful
measurements of the C/O ratio in K-type bulge giants find somewhat lower ratios
than this. For example, \citet{Mel08} find a mean of 0.24 in a sample of 19
K-type giants, and \citet{Ryde10} find a mean of 0.26 among 11 stars. Although
the Spitzer sample has a somewhat higher mean C/O ratio than these K giant
samples, we do not think that the sample contains stars that underwent 3DUP.
Rather, there could be systematic effects between the analysis of the AGB stars
here and the hotter K-type giants.

\begin{figure}
  \centering
  \includegraphics[width=\columnwidth,bb=84 370 536 698, clip]{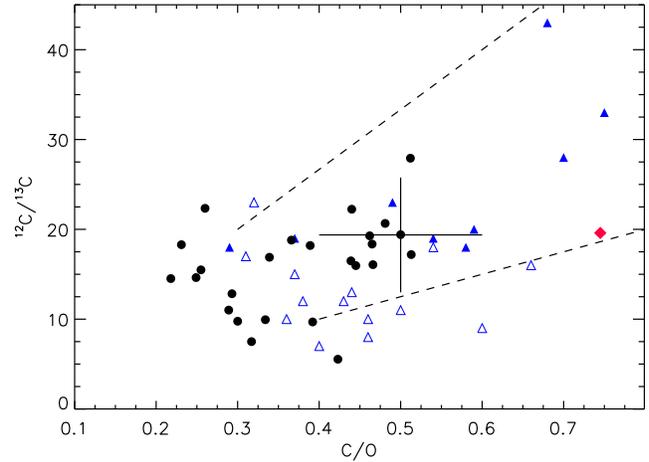}
  \caption{Carbon isotopic ratio vs.\ C/O. Filled black circles represent stars
    from this work, while filled blue triangles are Tc-rich MS- or S-type, open
    blue triangles Tc-poor M-type giants from \citet{SL90}. The filled red
    diamond symbol is the Tc-rich star Plaut 3-942. The two dashed lines
    indicate the evolution of atmospheres with (C/O,$^{12}$C/$^{13}$C)=(0.30,20)
    and (0.40,10) by adding $^{12}$C to them. A typical error bar is shown for
    one star (J174917.0-293502).}
  \label{12C13C_CO}
\end{figure}

Stars that have undergone 3DUP events may be revealed by a plot of the carbon
isotopic ratio vs.\ C/O, since both these ratios are expected to be enhanced.
Such a diagram is presented in Fig.~\ref{12C13C_CO}. Included in that diagram
are solar neighbourhood red giants from \citet{SL90}: Filled blue triangles
represent giants of type M or MS with Tc, open blue triangles are M-type giants
without Tc. Tc-poor MS/S-type stars are not included here because they are
thought to be the product of binary mass transfer. To guide the eye, the
evolution of AGB atmospheric abundance patterns by adding $^{12}$C to them is
indicated by the two dashed lines in Fig.~\ref{12C13C_CO}. This is a rather
simplistic view because it ignores possible effects of deep mixing on the AGB
\citep{Bus10}. The operation of deep mixing can considerably mask the 3DUP
evidence because $^{12}$C will be consumed to from $^{14}$N. Indeed, not all of
the Tc-rich stars in Fig.~\ref{12C13C_CO} have an enhanced C/O or
$^{12}$C/$^{13}$C ratio, and there are Tc-poor stars that have a higher C/O ratio
than some of the Tc-rich stars have. The disc giants discussed in the Appendix
may give a further indication of typical abundance ratios found in pre-3DUP
giants. However, there are a few disc stars in Fig.~\ref{12C13C_CO} that have
obviously added a considerable amount of $^{12}$C to their atmospheres. The only
star from our sample that is displaced from the rest is Plaut~3-942 (filled red
diamond symbol), a star with known Tc content. Its $^{12}$C/$^{13}$C ratio is
uncertain, but seems to be higher than that of most of the other sample stars.
Unfortunately, it was not possible to determine the $^{12}$C/$^{13}$C ratio in
Plaut 3-626.

The Spitzer sample star with the highest ratios is J180308.7-295220 at C/O=0.51
and $^{12}$C/$^{13}$C=28. Note, however, that the errors in these quantities are
correlated, an overestimate in C/O also leads to an overestimate in
$^{12}$C/$^{13}$C, see Sect.~\ref{err_est}. This star does not show an enhanced
Y abundance in its atmosphere (Table~\ref{tab_results}). Also, none of the cool
stars for which no metallicity or C/O ratio could be determined has a
significantly enhanced $^{12}$C/$^{13}$C. Therefore, we believe that none of the
Spitzer sample stars has undergone a 3DUP event yet (although they may have
undergone He-shell flashes without 3DUP). The same conclusion is reached from
the location of the sample stars in a dust mass-loss rate ($K-[22]$) vs.\
pulsation period diagram, which allows to distinguish between Miras with and
without Tc \citep{Utt13}. This also implies that C/O and $^{12}$C/$^{13}$C are
not useful as ``evolutionary clock'' for the Spitzer sample, as initially
anticipated for the observations.

\begin{figure}
  \centering
  \includegraphics[width=\columnwidth,bb=84 370 536 698, clip]{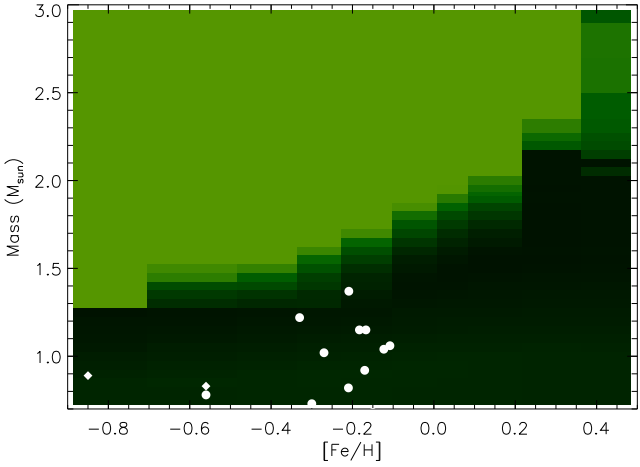}
  \caption{Formation of carbon stars and 3DUP behaviour in the initial mass vs.\
    metallicity plane of models from \citet{Mar13}. Light coloured areas
    indicate a combination of mass and metallicity where the model star becomes
    carbon-rich for at least a short while on the AGB. The darker the colour,
    the lower the maximum C/O ratio reached by the model star. The white symbols
    are sample stars for which a current pulsation mass and a metallicity was
    determined: circles are for Spitzer sample stars, diamonds for Plaut stars.}
  \label{mm}
\end{figure}

With the sample of stars for which we have information on the current mass and
the metallicity, we may put constraints on AGB evolutionary models that make
predictions concerning the occurrence of 3DUP and the formation of carbon stars.
It is difficult to obtain the mass {\it and} metallicity of an AGB star, because
it requires a star to pulsate in order to derive its mass, but dynamic effects
caused by the pulsations hamper the determination of the metallicity.
Nevertheless, in Fig.~\ref{mm} the stars are plotted in the mass -- metallicity
plane, along with predictions from model tracks calculated with the COLIBRI
code \citep{Mar13}. The maximum C/O ratio reached by the model stars is
colour-coded in that diagram: stars in the light coloured area in that plane
turn into carbon stars at least for some time on the AGB, the darker the colour
the lower the maximum C/O ratio. The minimum initial mass required for a star to
experience 3DUP or to become a carbon star increases with increasing
metallicity because the mixing is less efficient at higher metallicity. At solar
metallicity, the models of \citet{Mar13} predict that a minimum initial mass of
1.90\,$M_{\sun}$ is required to form a carbon star, and the interpulse luminosity
these stars is predicted to be in the range  $2000 - 6000 L_{\sun}$. Our sample
stars (white symbols) have sub-solar metallicity, but their current masses would
be too small to undergo significant 3DUP.

The two Plaut stars, which are at moderately low metallicity and clearly
experienced 3DUP, are also in a region of the diagram where the models do not
predict 3DUP to happen, clearly in contradiction with observations. In
principle, 3DUP should be eased in these stars because of their lower
metallicity. Nevertheless, at the metallicity of the Plaut stars, the models
predict that a higher initial mass is required for 3DUP to occur than the Plaut
stars have. The initial masses of the stars must have been higher than the
current masses, but it is not expected that they have lost a significant amount
of their mass. The largest mass loss occurs during the super-wind phase on the
AGB, which our stars have not yet entered, except maybe the ones with the
highest mass-loss rates. If the determined current masses are reliable and if
they have not been lowered much from the initial value by mass loss, then the
evolutionary models would predict 3DUP to occur at too high initial masses at
these moderately low metallicities.

A number of sample stars with a high current mass but without measured
metallicity might fall in the range where carbon star formation is predicted
by the models. Since all of these stars are oxygen-rich, they are problematic
for the evolutionary models, unless they have a considerably higher metallicity
than the other sample stars. This may be the case for the LAVs in the Spitzer
subsample, see Sect.~\ref{sec_mass}. We caution that there is also the
possibility that the linear pulsation models are not adequate to derive the
current masses of some of the stars, or that the input stellar parameters for
the mass determination (e.g.\ temperature) are in error.

The lower panel of Fig.~\ref{FeH_CO} shows the C/O ratio as a function of
metallicity. It is clear that the Spitzer sample stars are quite homogeneous in
their overall composition, with metallicities mostly sub-solar and all C/O
ratios below the solar value. The Plaut stars (red symbols) are clearly offset
from this group. Our sample does however not allow to infer the evolution of the
C/O ratio with metallicity in the bulge region. For models and observations of
the C/O evolution, we refer to \citet{Ces09}.

\subsection{Metal abundances}

\begin{figure}
  \centering
  \includegraphics[width=\columnwidth,bb=71 406 430 509, clip]{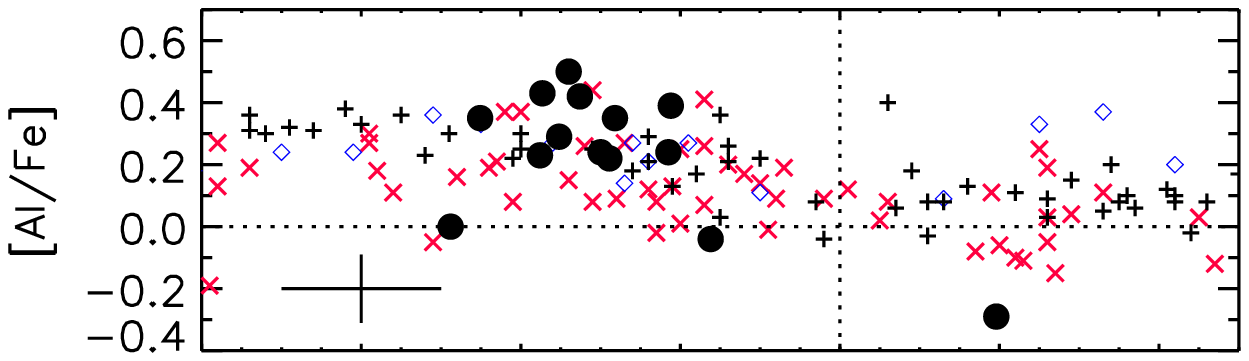}
  \includegraphics[width=\columnwidth,bb=71 406 430 509, clip]{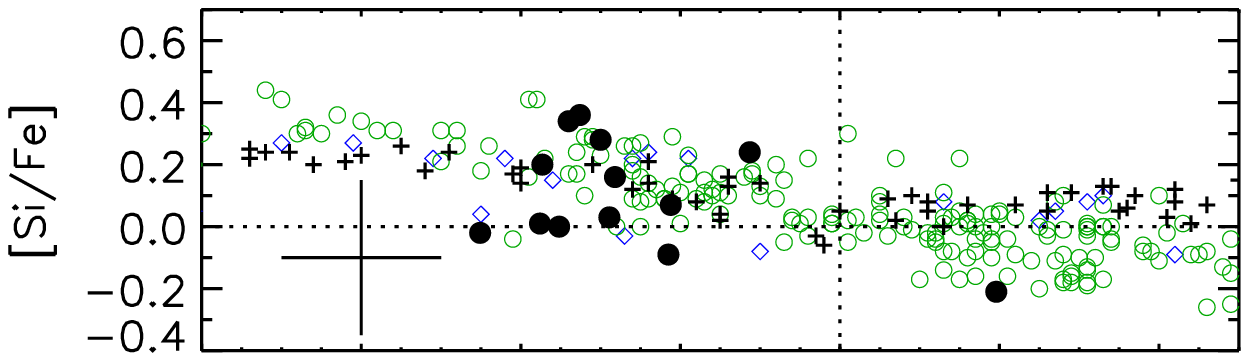}
  \includegraphics[width=\columnwidth,bb=71 406 430 509, clip]{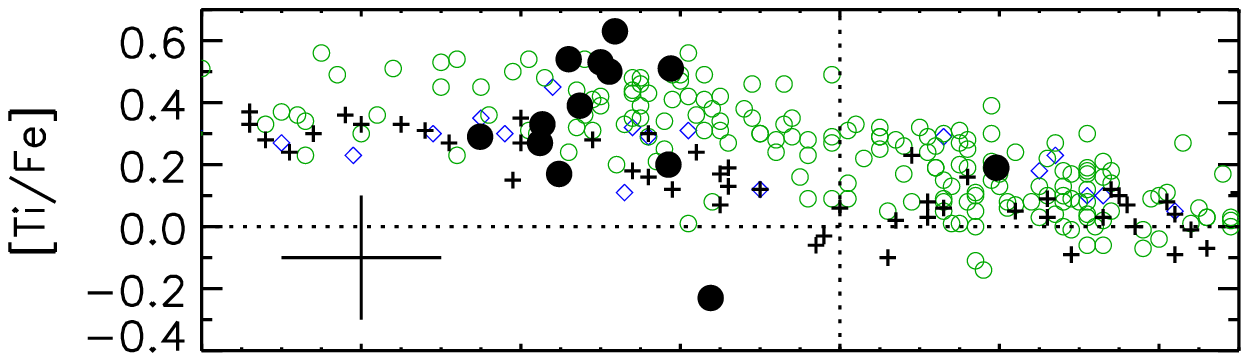}
  \includegraphics[width=\columnwidth,bb=71 369 430 509, clip]{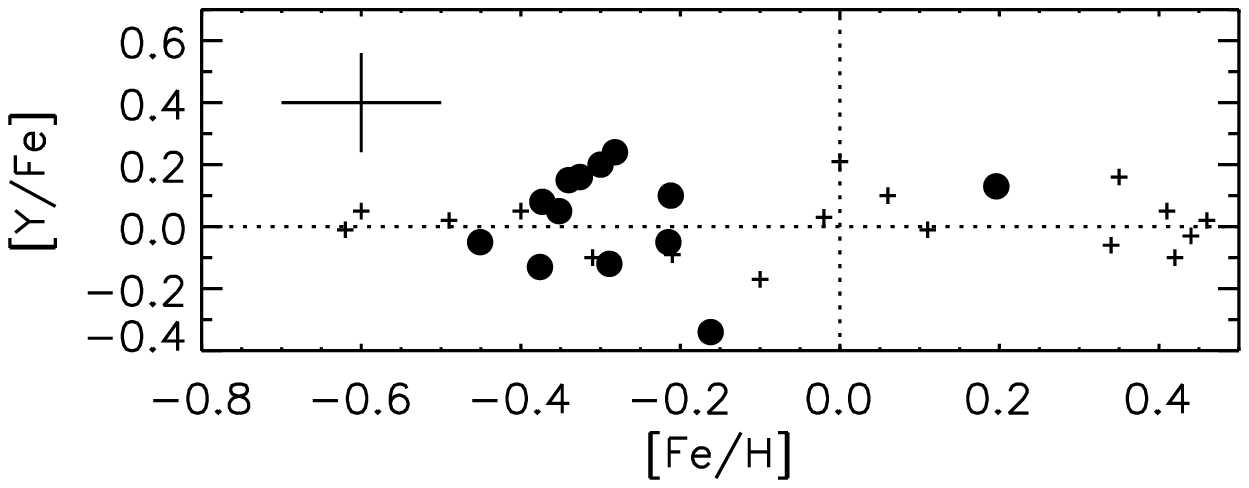}
  \caption{Abundances of [Al/Fe], [Si/Fe], [Ti/Fe], and [Y/Fe] as a function of
    [Fe/H] (from top to bottom) in samples of Galactic bulge stars. Filled black
    circles are data points from this work, black plus symbols are dwarf and
    subgiant stars from \citet{Bens13}, blue open diamonds are from Baade's
    field from \citet{Alv10}, green open circles from
    \citet[][only Baade's field]{Gon11}, and red crosses are from Plaut's field
    in the outer bulge from \citet{Joh12}. The big symbols on the left hand side
    of each panel show typical error bars.}
  \label{XM_MH}
\end{figure}

Only the Spitzer sample stars were observed in the additional K-band setting;
the individual metal abundances derived from them are discussed here. The
results of these measurements are presented in Fig.~\ref{XM_MH}. The scatter in
the abundances may be fully explained by measurement uncertainty (see the error
bars on the left hand side of the panels). The abundances are compared primarily
to measurements in similar regions of the bulge as the location of the Spitzer
sample stars \citep{Bens13,Alv10,Gon11}. Only for Al measurements from the outer
bulge (Plaut's field, $b\approx-8\degr$) from \citet{Joh12} are included. The
Spitzer sample is too small and spans too small a range in metallicity to probe
for any trends of the elements with metallicity. However, the abundances are
suited to check if the AGB stars in the present sample descend from the same
populations that can be found among less evolved bulge stars. A look at
Fig.~\ref{XM_MH} can answer this question with ``yes'': The relative abundances
of Al, Si, Ti, and Y agree well with the range of abundances that are seen in
other bulge stars at these metallicities. The fact that Y, one of the light
s-process elements, is not significantly enhanced also argues against 3DUP being
active in the Spitzer sample stars for which the Y abundance could be derived.
Also, no correlation between [Y/Fe] and C/O is found in our data. C and Y are
thus probably not produced in situ and the abundances likely reflect the
primordial values.

\subsection{Radial velocities}

Here only the heliocentric radial velocities of the 37 Spitzer sample stars
(column~2 in Table~\ref{tab_results}) are discussed. For the radial velocity
dispersion of AGB stars in the outer bulge, see \citet[][Section 4.7.3]{Utt12}.

\begin{figure}
  \centering
  \includegraphics[width=\columnwidth,bb=68 370 540 700, clip]{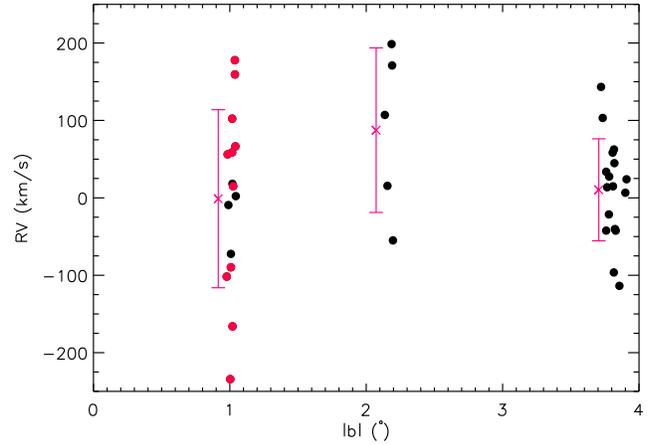}
  \caption{Radial velocity vs.\ absolute value of Galactic latitude for the
    Spitzer sample stars. Red symbols represent stars located north of the
    Galactic plane, at positive latitude. The large symbols next to the three
    latitude groups show the mean and standard deviation of the radial velocity
    of the respective group.}
  \label{RV_b}
\end{figure}

The radial velocities of the Spitzer stars have a mean of
$\left<{\rm RV}\right>=16.1\pm15.8$\,km\,s$^{-1}$ and a dispersion of
$\sigma_{\rm RV}=95.8\pm11.3$\,km\,s$^{-1}$. However, it appears that the RV
dispersion depends on the latitude of the fields (Fig.~\ref{RV_b}): while in
the field at $b\sim-3.8\degr$ the dispersion of the 17 stars is only
$65.9\pm11.6$\,km\,s$^{-1}$, the fields at $b\sim\pm1\degr$ (15 stars) have a
combined dispersion of $115.0\pm21.7$\,km\,s$^{-1}$. It has been shown that the
metal-rich stellar population of the Galactic bulge has a radial velocity
dispersion that significantly increases when approaching the plane, whereas the
metal-poor population has a constant $\sigma_{\rm RV}$ of
90 -- 100\,km\,s$^{-1}$ throughout the bulge \citep{Bab10,Utt12,Bab14}. This
increasing radial velocity dispersion is a result of the action of the Galactic
bar. We thus suggest that the Spitzer sample also follows a bar-like kinematic
pattern, and that it belongs to the metal-rich population. The metallicities
measured in the present paper are in agreement with this picture. Note that Mira
stars in the bulge are actually found to trace a tilted bar \citep{GB05}. As
shown above that the Plaut stars are more metal-poor. The radial velocity
dispersion of AGB stars in the outer bulge suggests that they descend from the
metal-poor, spheroidal component of the bulge.

\section{Conclusions}\label{conclusio}

We present an analysis of high-resolution near-IR spectra of a sample of
Galactic bulge AGB stars. The sample consists of a larger sub-sample from the
inner and intermediate bulge (Spitzer sample) and a smaller sample from the
outer bulge (Plaut stars). Using state-of-the-art hydrostatic model atmospheres
for cool stars and the best available atomic and molecular line lists for the
observed wavelength range, we are able to successfully model the spectra of
stars down to $T_{\rm eff}\approx3100$\,K. General metallicities [Fe/H], the
carbon-to-oxygen ratio C/O, carbon isotopic ratio $^{12}$C/$^{13}$C, the
abundance of Si, Al, Ti, and Y, and heliocentric radial velocities are
determined. The abundance analysis relies on the assumption of an
$[\alpha/{\rm Fe}]$ trend with [Fe/H] because of the low sensitivity of OH lines
to the O-abundance in cool, M-type giants. We caution that calculation of both
the model atmosphere and the spectral synthesis with consistent abundances is
required in such cool stars for a reliable modelling. Current stellar masses are
also derived from linear pulsation models.

This paper touches on the problem of the apparent lack of intrinsic carbon
stars in the Galactic bulge. All sample stars are oxygen-rich, there is no
carbon star among them. The C/O and carbon isotopic ratios measured in the
sample stars suggest that the Spitzer sample stars did not undergo the third
dredge-up. They appear to have a composition typical for post-first dredge-up
red giants. Furthermore, no increased CO band head strength was found among the
stars for which we could not measure the overall composition, excluding the
known technetium-rich stars. Note, however, that the effect of 3DUP on the
carbon abundance may be considerably masked by the operation of deep mixing on
the AGB. The conclusion of a lack of 3DUP in these stars is supported by the
fact that Y is not enhanced in the stars with observations of a line of that
s-element, and the position of the Spitzer stars in a dust mass-loss rate vs.\
period diagram \citep{Utt13}. On the other hand, two of the Plaut stars have a
clearly enhanced C/O ratio, as expected from the fact that Tc lines have been
found earlier in their optical spectra.

However, detailed AGB evolutionary models, e.g.\ \citet{Mar13}, are in conflict
with our measurements of metallicities, masses, and 3DUP occurrence in the
sample stars. Those models predict no 3DUP for stars with initial masses as low
as the current masses of some of the Tc-rich Plaut stars, but do predict 3DUP
and carbon star formation for the more massive stars in our sample. Though it
is impossible to determine the metallicity and C/O ratio for the coolest and
putatively most massive stars in the sample, it is clear that they are not
carbon-rich. Assuming a similar metallicity distribution as the other Spitzer
stars, one would expect from the models that they evolve into carbon stars. The
reason for this could be that the masses estimated with our linear
pulsation models are not accurate enough because of uncertainties on the input
parameters and/or invalid assumptions for the models. We suggest that 3DUP is
found in the Plaut stars but not in the Spitzer stars because of their lower
metallicity, see below. On the other hand, the C-star lifetime is very short at
high metallicity, and also a high He abundance could prevent 3DUP and C-star
formation \citep{Kar14}.

It has been suggested that a high oxygen abundance could prevent the formation
of carbon stars in the Galactic bulge \citep{Fea07}.
However, the current consensus is that the O-abundance trend in bulge stars
is very similar to that of local thick disc stars \citep{Mel08,Ryde10,Bens13}.
We suggest that at the slightly subsolar metallicity of the Spitzer stars,
carbon star formation is suppressed because 3DUP is suppressed altogether,
either because of the high metallicity or possibly because of a high He
abundance. Hence, an enhanced oxygen abundance is not required to explain the
lack of carbon stars in the Galactic bulge. 3DUP does occur at the lower
metallicity of the Plaut stars in the outer bulge, but there may not be
sufficient 3DUP events to turn them into C-rich stars, and/or because of the
action of deep mixing. From the masses that we derive for these stars one might
expect C-star formation, but note the uncertainties in the mass determinations.

The Spitzer sample stars have on average slightly sub-solar metallicity but
there are few that have super-solar metallicity. This is in agreement with the
results of previous observations of cool giant samples in the Galactic bulge,
though the comparison on the level of individual stars is more uncertain. The
lack of super-solar metallicity stars could be caused by metallicity-dependent
stellar winds \citep{Coh08}.

The Plaut stars are found to be more metal-poor, in agreement with earlier
suggestions by \citet{Utt12} that were based on kinematical arguments. Thus, we
suggest that the AGB stars in the outer Galactic bulge mainly descend from the
metal-poor bulge population ($[{\rm Fe}/{\rm H}]\sim-0.6$), not from the
metal-rich one ($[{\rm Fe}/{\rm H}]\sim+0.3$), although about 30\% of the RGB
stars in a field at $(l,b)=(0\degr,-10\degr)$ belong to the metal-rich bulge
population \citep{Utt12}. The metallicity inferred for Galactic bulge PNe
\citep{Bue13} is higher than what we find among AGB stars. It could be that
metal-rich stars are more likely to form PNe, which would be in agreement with
the call for metallicity-dependent stellar winds.

The abundances of Al, Si, Ti, and Y measured for some of the Spitzer sample
stars suggest that they descend from the bulk of the population in their bulge
region. The radial velocity dispersion of the Spitzer stars suggests that they
follow a bar-like kinematic pattern, characteristic for the metal-rich bulge
population. The Plaut stars, on the other hand, seem to belong to the
metal-poor, spheroidal bulge component.

Finally, the masses of the sample stars suggest that the Galactic bulge is not
exclusively old, a younger and more massive population ($\sim1.3M_{\sun}$) seems
to be present. This would be in agreement with age estimates of metal-rich dwarf
stars in the Galactic bulge \citep{Bens13}.

\section*{Acknowledgments}
We thank Katia Cunha and Livia Origlia for kindly providing their spectra of the
comparison stars, as well as Paola Marigo for providing model tracks from the
COLIBRI code. SU acknowledges support from the Austrian Science Fund (FWF) under
project P~22911-N16 and from the Fund for Scientific Research of Flanders (FWO)
under grant number G.0470.07, TL acknowledges support from the FWF under
projects P~23737-N16 and P~21988-N16. BA acknowledges the support from the
{\em project STARKEY} funded by the ERC Consolidator Grant, G.A.\ n.~615604. NR
is a Royal Swedish Academy of Sciences Research Fellow supported by a grant from
the Knut and Alice Wallenberg Foundation. NR acknowledges support from the
Swedish Research Council, VR and by Funds from Kungl.\ Fysiografiska
S\"allskapet i Lund. This publication is based on observations at the Very Large
Telescope of the European Southern Observatory, Cerro Paranal/Chile under
Programmes 081.D-0669(A) and 383.D-0685(A). This publication makes use of data
products from the Two Micron All Sky Survey, which is a joint project of the
University of Massachusetts and the Infrared Processing and Analysis
Center/California Institute of Technology, funded by the National Aeronautics
and Space Administration and the National Science Foundation, as well as of the
Wide-field Infrared Survey Explorer, which is a joint project of the University
of California, Los Angeles, and the Jet Propulsion Laboratory/California
Institute of Technology, funded by the National Aeronautics and Space
Administration. This research has made use of the VizieR catalogue access tool,
CDS, Strasbourg, France. The original description of the VizieR service was
published in A\&AS 143, 23.



\appendix
\label{append}

\section{Abundances in five field red giants}\label{field_giants}
Observations of five red giant stars in the field have been obtained in the
same CRIRES run. These spectra were also analysed and the results are reported
here for the sake of completeness.

The determination of the effective surface temperature and gravity of the field
stars relied purely on literature values and near-IR photometry. Temperatures of
the stars are given by \citet{Wri03}, which are based on spectral type
determinations. As a second estimate, we applied the $T_{\rm eff} - (J-K)_0$
calibration of \citet{LU12} to the $(J-K)_0$ colour that was derived from
2MASS photometry \citep{2MASS} and the interstellar reddening map of
\citet{Schleg}. A compromise value between these two estimates was adopted for
the model atmospheres. The surface gravity $\log g$ was estimated also with the
$\log g - (J-K)_0$ calibration of \citet{LU12}. The stellar parameters of the
field giants are summarised in Table~\ref{tab_a1}.

\begin{table*}
\centering
\begin{minipage}{164mm}
\caption{Stellar parameters of the field red giants. Meaning of the columns:
object name; range of spectral types found in VizieR; J, H, and K magnitudes
from 2MASS; interstellar extinction in the V-band from \citet{Schleg};
de-reddened $J-K$ colour; effective temperature derived from the
$T_{\rm eff}-(J-K)_0$ calibration of \citet[][L12]{LU12}; effective temperature
listed in \citet[][W03]{Wri03}; adopted effective temperature; adopted $\log g$
of the model atmosphere, estimated with the $\log g - (J-K)_0$ calibration of
\citet{LU12}.}
\label{tab_a1}
\begin{tabular}{llccccccccc}
\hline
Object name       & Sp.\ type  & $J$   & $H$   & $K$   & $A_{\rm V}$ & $(J-K)_0$ & $T_{\rm eff}$ & $T_{\rm eff}$ & $T_{\rm eff}$  & $\log g$ \\
                  &            & (mag) & (mag) & (mag) & (mag)     &           & (K, L12)    & (K, W03)    & (K, adopted) & (cgs)    \\
\hline
HD136546 & K5\,III    & 4.445 & 3.515 & 3.313 & 0.730     & 1.021     &  3785       & 3950        & 3900         & 1.20 \\
HD140672 & K5-M1\,III & 4.207 & 3.251 & 3.021 & 0.947     & 1.042     &  3740       & 3720        & 3750         & 1.10 \\
HD141189 & M0-2\,III  & 4.542 & 3.551 & 3.252 & 0.407     & 1.228     &  3295       & 3620        & 3400         & 0.38 \\
HD141311 & K5-M1\,III & 4.849 & 3.862 & 3.701 & 0.776     & 1.030     &  3766       & 3720        & 3750         & 1.10 \\
HD141938 & M0-2\,III  & 4.314 & 3.367 & 3.133 & 0.574     & 1.094     &  3624       & 3720        & 3650         & 0.90 \\
\hline
\end{tabular}
\end{minipage}
\end{table*}

The CRIRES spectra of the field stars, which are of high quality, were analysed
in the same way as the bulge AGB stars in the main part of the paper, except
that for the $\alpha$-abundance a more disk-like relation was adopted, with
$[\alpha/{\rm Fe}]=+0.4$ up to $[{\rm Fe}/{\rm H}]=-1.0$, followed by a decline 
of 0.4\,dex/dex for higher metallicities. The results of the analysis are
summarised in Table~\ref{tab_a2}. All five stars have close to solar
metallicity, and slightly sub-solar C/O ratios \citep[on the scale of][]{Caf08}.

\begin{table*}
\centering
\begin{minipage}{120mm}
\caption{Radial velocities and abundances measured for the field red giants.
Meaning of the columns: object name; heliocentric radial velocity; metallicity;
C/O ratio; carbon isotopic ratio, and relative abundances of Al, Si, Ti, and Y.}
\label{tab_a2}
\begin{tabular}{lrrrrrrrr}
\hline
Object name       & RV$_{\rm helio}$ & [Fe/H]  & C/O & $^{12}$C/$^{13}$C & [Al/Fe]  & [Si/Fe]  & [Ti/Fe]   & [Y/Fe]   \\
                  & (km\,s$^{-1}$) &         &     &                 &         &         &          &         \\
\hline
HD136546 &  $-9.7$       & $+0.05$ & 0.38 & 15.4            & $+0.11$ & $+0.05$ & $+0.04$ & $-0.18$ \\
HD140672 & $-27.3$       & $+0.04$ & 0.43 & 15.6            & $+0.20$ & $-0.02$ & $+0.12$ & $-0.20$ \\
HD141189 & $-97.3$       & $-0.09$ & 0.28 & 14.3            & $+0.03$ & $-0.07$ & $+0.14$ & $-0.28$ \\
HD141311 &  $64.6$       & $-0.23$ & 0.50 & 12.3            & $+0.15$ & $+0.06$ & $+0.27$ & $-0.17$ \\
HD141938 & $-38.7$       & $-0.05$ & 0.30 & 14.4            & $+0.05$ & $+0.08$ & $+0.02$ & $-0.08$ \\
\hline
\end{tabular}
\end{minipage}
\end{table*}

\end{document}